\documentclass{jfm}

\title{Rossby Number Effects on Columnar Eddy Formation and the Energy
Dissipation Law in Homogeneous Rotating Turbulence}

\author{T. Pestana\aff{1}
  \corresp{\email{t.pestana@tudelft.nl}}
 \and S. Hickel\aff{1}}

\affiliation{\aff{1}Department for Aerodynamics, Faculty of Aerospace
Engineering, Kluyverweg 2, 2629 HS Delft, The Netherlands}

\usepackage{graphicx}
\usepackage{psfrag}  \usepackage{color}
\usepackage{tensor}
\usepackage{listings}
\usepackage{mathrsfs}
\usepackage{amsmath}
\usepackage{calc}
\usepackage{keyval}
\usepackage{xcolor,color}
\usepackage{hyperref}
\usepackage[position=top]{subfig}
\usepackage{textcomp}
\usepackage{filecontents}
\usepackage{tabularx,blkarray}
\usepackage{enumerate}
    \hypersetup{
        unicode=false,          pdftoolbar=true,        pdfmenubar=true,        pdffitwindow=false,     pdfstartview={FitH},    pdfcreator={Creator},   pdfkeywords={keyword1, key2, key3}, pdfnewwindow=true,      colorlinks=true,        linkcolor=black,        citecolor=black,        filecolor=magenta,      urlcolor=cyan           }
\usepackage{cleveref}
    \crefname{section}{section}{sections}
    \Crefname{section}{Section}{Sections}
    \crefname{subsection}{section}{Sections}
    \Crefname{subsection}{Section}{Sections}
    \crefname{figure}{figure}{figures}
    \Crefname{figure}{Figure}{Figures}
    \crefname{equation}{equation}{equations}
    \Crefname{equation}{Equation}{Equations}
    \crefname{table}{table}{tables}
    \Crefname{table}{Table}{Tables}

\usepackage{tikz}
\usetikzlibrary{decorations.pathreplacing}
\usetikzlibrary{arrows.meta}

\usepackage{soul}

\definecolor{myorange}{rgb}{0.93, 0.69, 0.13}
\definecolor{mygray}{rgb}{0.5, 0.5, 0.5}

\definecolor{run01}{rgb}{0.000000, 0.000000, 1.000000}
\definecolor{run02}{rgb}{1.000000, 0.000000, 0.000000}
\definecolor{run03}{rgb}{0.000000, 1.000000, 0.000000}
\definecolor{run04}{rgb}{0.000000, 0.000000, 0.172414}
\definecolor{run05}{rgb}{1.000000, 0.103448, 0.724138}
\definecolor{run06}{rgb}{1.000000, 0.827586, 0.000000}
\definecolor{run07}{rgb}{0.000000, 0.344828, 0.000000}
\definecolor{run08}{rgb}{0.517241, 0.517241, 1.000000}
\definecolor{run09}{rgb}{0.620690, 0.310345, 0.275862}
\definecolor{run10}{rgb}{0.000000, 1.000000, 0.758621}
\definecolor{run11}{rgb}{0.000000, 0.517241, 0.586207}
\definecolor{run12}{rgb}{0.000000, 0.000000, 0.482759}
\definecolor{run13}{rgb}{0.586207, 0.827586, 0.310345}
\definecolor{run14}{rgb}{0.965517, 0.620690, 0.862069}
\definecolor{run15}{rgb}{0.827586, 0.068966, 1.000000}
\definecolor{run16}{rgb}{0.482759, 0.103448, 0.413793}
\definecolor{run17}{rgb}{0.965517, 0.068966, 0.379310}
\definecolor{run18}{rgb}{1.000000, 0.758621, 0.517241}
\definecolor{run19}{rgb}{0.137931, 0.137931, 0.034483}
\definecolor{run20}{rgb}{0.551724, 0.655172, 0.482759}
\definecolor{run21}{rgb}{0.965517, 0.517241, 0.034483}

\DeclareRobustCommand\straightline{\tikz[baseline=-0.6ex]\draw[-,line
width=1mm] (-1ex,.5ex)--++(.3cm,0);}

\newcommand{\oblack}{
\tikz[baseline]{
\node[draw,scale=0.5,circle,line width=.35mm]
() at (.25cm,.6ex){};
}~}

\newcommand{\blackline}{
\tikz[baseline]{
\draw[-,line width=.35mm,black] (-1ex,.5ex)--++(.3cm,0);
}~}

\newcommand{\blueline}{
\tikz[baseline]{
\draw[-,line width=.35mm,blue] (0,.5ex)--++(.3cm,0);
}~}

\newcommand{\blackdashed}{
\tikz[baseline]{
\draw[-,line width=.35mm,dashed] (0,.5ex)--++(.5cm,0);
}~}

 \renewcommand{\v}[1]{\ensuremath{\mathbf{#1}}} \newcommand{\gv}[1]{\ensuremath{\boldsymbol{#1}}} \newcommand{\norm}[1]{\left\lVert#1\right\rVert}
\newcommand{\uv}[1]{\ensuremath{\mathbf{\hat{#1}}}}

\newcommand{\mean}[1]{\langle #1 \rangle}
\newcommand{\boxmean}[1]{\left\langle #1 \right\rangle_\mathcal{L}}

 \newcommand{\pd}[2]{\frac{\partial #1}{\partial #2}}

\newcommand{\pdinline}[2]{{\partial{#1}}/{\partial{#2}}}
\newcommand{\dinline}[2]{{\mbox{d}{#1}}/{\mbox{d}{#2}}}

\newcommand{\diff}{\mathop{}\!\mathrm{d}}

\newcommand{\mypara}{{\mkern3mu\protect\vphantom
                     {\perp}\vrule depth 0pt\mkern2mu\vrule depth 0pt\mkern3mu}}

\newcommand{\urms}{u'}
\newcommand{\ui}{u_i}

\newcommand{\vort}{\gv{\omega}}
\newcommand{\dissip}{\varepsilon}

\newcommand{\dissipnu}{\dissip_{\nu}}
\newcommand{\dissipI}{\dissip_{I}}
\newcommand{\kenfrisch}{K}
\newcommand{\ken}{\kenfrisch}

\newcommand{\rpp}{r_\perp}
\newcommand{\rpar}{r_\mypara}

\newcommand{\dpp}{\mathcal{L}_\perp}
\newcommand{\dpar}{\mathcal{L}_\mypara}
\newcommand{\wn}{\kappa}

\newcommand{\wndpp}{\wn_f \mathcal{L}_\perp}
\newcommand{\wndpar}{\wn_f \mathcal{L}_\mypara}
\newcommand{\wnlpp}{\wn_f \ell_\perp}
\newcommand{\wnlpar}{\wn_f \ell_\mypara}

\newcommand{\kzeman}{\wn_{\Omega}}
\newcommand{\wnf}{\wn_f}
\newcommand{\urmsiso}{\urms^{\,\text{iso}}}
\newcommand{\lscale}{l_0}
\newcommand{\lscalepar}{\ell_{0\,\mypara}}
\newcommand{\lscalepp}{\ell_{0\,\perp}}
\newcommand{\lf}{\ell_f}
\newcommand{\tf}{\tau_f}
\newcommand{\ts}{\tau_s}
\newcommand{\teddy}{T_e}
\newcommand{\trelax}{\tau_3}
\newcommand{\tnl}{\tau_{nl}}
\newcommand{\tnliso}{\tnl^{\,\text{iso}}}
\newcommand{\lkolmo}{\eta}
\newcommand{\tomg}{\tau_\Omega}

\newcommand{\uf}{u_f}
\newcommand{\uscale}{u_0}

\newcommand{\lpp}{\ell_\perp}
\newcommand{\lpar}{\ell_\mypara}
\newcommand{\lpariso}{\lpar^{\,\text{iso}}}
\newcommand{\lppiso}{\lpp^{\,\text{iso}}}
\newcommand{\liso}{\ell^{\,\text{iso}}}

\newcommand{\ceps}{C_\varepsilon}
\newcommand{\cepsiso}{\ceps^{\,\text{iso}}}

\newcommand{\rossby}{Ro}
\newcommand{\reynoldslmbda}{Re_{\lambda}}
\newcommand{\reynoldseps}{Re_{\varepsilon}}

\newcommand{\rossbyeps}{Ro_{\varepsilon}}
\newcommand{\rossbylmbda}{Ro_{\lambda}}

\usepackage{etoolbox}
\makeatletter
\patchcmd{\NAT@test}{\else \NAT@nm}{\else \NAT@hyper@{\NAT@nm}}{}{}
\makeatother
\begin{document}

\maketitle

\begin{abstract}
Two aspects of homogeneous rotating turbulence are quantified through forced
Direct Numerical Simulations in an elongated domain, which is in the direction
of rotation about $340$ times larger than the typical initial eddy size. First,
by following the time evolution of the integral length-scale along the axis of
rotation $\lpar$, the growth rate of the columnar eddies and its dependency on
the Rossby number $\rossbyeps$ is determined as $\gamma =4
\exp(-17\rossbyeps)$, where $\gamma$ is the non-dimensional growth rate. Second,
a scaling law for the energy dissipation rate $\dissipnu$ is sought. Comparison
with current available scaling laws shows that the relation proposed by
\citet{baqui:2015}, i.e., $\dissipnu\sim \urms^3/\lpar$, where $\urms$ is the
r.m.s. velocity, approximates well part of our data, more specifically the range
$0.39\le\rossbyeps\le1.54$. However, relations proposed in the literature fail
to model the data for the second and most interesting range, i.e.,
$0.06\le\rossbyeps\le0.31$, which is marked by the formation of columnar eddies.
To find a similarity relation for the latter, we exploit the concept of a
spectral transfer time introduced by \citet{kraichnan:1965}. Within this
framework, the energy dissipation rate is considered to depend on both the
non-linear time-scale and the relaxation time-scale. Thus, by analyzing our
data, expressions for these different time-scales are obtained that results
in~$\dissipnu\sim\urms^4/(\lpp^2\rossbyeps^{0.62} \tnliso)$, where $\lpp$ is the
integral length-scale in the direction normal to the axis of rotation and
$\tnliso$ is the non-liner time-scale of the initial homogeneous isotropic
field.
\end{abstract}

\section{Introduction}

Many geophysical and man-made fluid flows are affected by the interaction
between system rotation and turbulence \citep{greenspan1968theory,Boffetta2012}.
Still, our knowledge of the effects of the Coriolis force on turbulence is
far from complete. An idealized approach to study rotating turbulence consists
in observing the evolution of an initial homogeneous isotropic flow in a
non-inertial rotating frame of reference. This way, early experimental studies
already revealed the main features of homogeneous rotating turbulence, although
a few of them did not meet the condition for homogeneity
\citep{Ibbetson1975,Hopfinger1982,Jacquin1990}. When the Rossby number
($\rossby$) was sufficiently small, i.e., the ratio of the rotational time scale
and the turbulent time scale, it was observed that the energy dissipation rate
$\dissipnu$ reduced with respect to the reference non-rotating isotropic case.
Further, the typical cloud of isotropic eddies found in isotropic flows was
strained, and grew in size to towards an array of flow structures aligned with
the axis of rotation (columnar eddies). These two features are the traits of
rotating turbulence and have been observed and analyzed in a number of recent
experimental and numerical investigations, see, e.g.,
\citet{Staplehurst2008,VanBokhoven2009,Mininni2009,Moisy2011,Delache2014,mininni:2012},
or \citet{Godeferd2015} for a review. Yet, it remains poorly understood how they
are quantitatively related.

For homogeneous isotropic turbulence, it is well accepted that the energy
dissipation rate scales as $\dissipnu \sim \uscale^3/\lscale$, where $\uscale$
and $\lscale$ are an integral velocity scale and an integral length scale,
respectively \citep{batchelor:book}. This relation can be interpreted on the
basis of phenomenological arguments as follows. Let us first assume that
$\dissipnu$ depends on a energy content, say $\uscale^2$, and on a time scale
$\ts$ characteristic of the downscale energy transfer: the spectral transfer
time. In homogeneous isotropic turbulence, the only time scale available to be
taken as $\ts$ is the time scale characteristic of the non-linear triadic
interactions, $\tnl$. If we further assume that $\tnl\sim\lscale/\uscale$, where
$\lscale$ is the typical size of the energy containing eddies, the dissipation
law for homogeneous isotropic turbulence can be recovered. But for systems in
which other time scales are also relevant, as is the case of
magnetohydrodynamics (MHD) or rotating turbulence, $\ts$ might be different from
$\tnl$. Within the context of MHD, \citet{kraichnan:1965} considered that $\ts$
is in fact composed of two time scales of opposing effects; the non-linear
time scale $\tnl$, which can also be considered as the measure of how fast
triple velocity correlations are built-up, and the decorrelation time scale
$\trelax$, which indicates how fast these correlations decay in time. Exploiting
these ideas, he suggested that the energy flux (energy dissipation rate) was
directly proportional to $\trelax$ and inversely proportional to $\tnl$.

Following this line of thought, one alternative to relate the energy dissipation
rate to the formation of columnar eddies in rotating turbulence is to find
approximations for $\tnl$ and $\trelax$ that involve integral length scales and
the rotation rate. However, this is not straightforward. First, owing to the
fact that the distribution of energy is not isotropic, two distinct integral
length scales in homogeneous rotating turbulence exist, i.e., $\lscalepp$ and
$\lscalepar$, which can be defined along the directions normal and parallel to
the axis of rotation, respectively. Which one then is relevant to form $\tnl$?
Second, how does $\trelax$ depend on the time scale imposed by the background
rotation, i.e., $\tomg=1/(2\Omega)$? In literature, a few dissipation laws for
homogeneous rotating turbulence have emerged from attempts to estimate the
energy flux \citep{zhou1995,Galtier2003,nazarenko:2011,baqui:2015}. Despite the
efforts to account for the effects of rotation, results available in current
literature regarding whether these laws generally hold or if they specifically
apply to a Rossby number range are inconclusive or even inconsistent.

Another problem, which is rather more technical, is the fact that the elongated
columnar flow structures restrict the maximum observation time in Direct
Numerical Simulations (DNS) of rotating turbulence. Because simulations of
homogeneous flows often consider periodic boundary conditions, a too small
domain size with respect to the characteristic size of the living eddies can
modulate the dynamics of the large scales and constrain their size. An obvious
solution to circumvent this problem and avoid numerical artifacts is either to
consider larger domains or to generate flow fields in which the characteristic
eddy size is smaller than the domain size. For example, in the DNS by
\citet{baqui:2015} the initial characteristic eddy size was $50$ times smaller
than the domain size. However, when $\rossby\ll1$ this may be still insufficient
and limit the simulation to a few eddy-turnover times.

In view of these shortcomings, this study addresses the two following
questions:\\

\begin{enumerate}[(i)]
\item \hspace{.25ex} What is the influence of the Rossby number in the growth
rate of the columnar eddies, in the absence of confinement effects?
\\[-.15cm]
\item \hspace{.25ex} Can we approximate the energy dissipation rate in
homogeneous rotating turbulence in a fashion similar to homogeneous isotropic
turbulence, i.e., in terms of a velocity scale, an integral length scale and the
rotation rate?\\
\end{enumerate}

For this purpose, we consider the evolution of an initial homogeneous isotropic
flow field in a rotating frame of reference. We conduct a systematic study that
consists of $21$ different rotation rates, thus covering a wide range of Rossby
numbers. Our DNS are carried out in an elongated computational domain that is
about $340$ times larger than the initial characteristic size of the flow
structures, provides enough room for the columnar eddies to grow freely. All
simulations are performed with a stochastic large-scale forcing that injects
energy at a constant rate. The forcing scheme is three-dimensional, isotropic,
and at all times uncorrelated with the velocity field. To the best of our
knowledge, the present database is unprecedented.

This work is organized as follows. In \cref{sec:methodology}, the governing
equations and the numerical method is detailed together with a description of
the simulations and their physical parameters. The influence of the
Rossby number in the growth rate of the columnar eddies is investigated in
\cref{sec:growthlpar}, and approximations for the energy dissipation rate are
finally offered in \cref{sec:dissiplaw}.

\section{Numerical Set-up}
\label{sec:methodology}

\subsection{Governing Equations and Numerical Method}
We consider an incompressible fluid in a triply periodic rectangular cuboid of
size $2\pi\mathcal{L}_1\times 2\pi\mathcal{L}_2 \times 2\pi\mathcal{L}_3$ that
rotates around $\gv{\Omega}$. Fluid motion is assumed governed by the
incompressible Navier-Stokes equations:
\begin{align}
     \nabla \cdot \v{u} &= 0, \label{eq:mass} \\
     \pd{\v{u}}{t} + (2\,\gv{\Omega} + \vort) \times \v{u} &= - \nabla q
     + \nu \nabla^2 \v{u} + \v{f}.
     \label{eq:ns}
\end{align}
Here, $\v{u}$, $\vort$ and $\v{f}$ are the velocity, the vorticity and an
external force, respectively. Time is denoted by $t$, the reduced pressure, into
which the centrifugal force is incorporated, is given by $q$, and $\nu$ denotes
the kinematic viscosity of the fluid. The rotation vector $\gv{\Omega}$ is
chosen to be aligned with the $3$-direction, i.e, $\gv{\Omega}=(0,0,\Omega)$,
where $\Omega$ is the rotation rate. The horizontal dimensions of the
rectangular cuboid (normal to the axis of rotation) are equal,
$\dpp=\mathcal{L}_1 = \mathcal{L}_2= 1$, whereas the vertical extension
(parallel to the axis of rotation) is by a factor of $8$ larger than the
horizontal dimensions, i.e. $\dpar=\mathcal{L}_3=8$.

The numerical method is essentially the same as in \citet{Pestana2019}.
\Cref{eq:mass,eq:ns} are solved by a de-aliased Fourier pseudo-spectral method
(2/3-rule), where the spatial gradients are computed with the aid of fast
Fourier transforms \citep{p3dfft}, and the time-stepper employs exact
integration of the viscous and Coriolis forces
\citep{Rogallo1977,Morinishi2001a} together with a third-order low-storage
Runge-Kutta scheme for the non-linear terms. The number of degrees of freedom is
$N_p=768^2\times6144$, which has been increased accordingly to the extended
domain size to resolve all scales of motion. The smallest and largest resolved
wavenumber per direction is $\wn_{min,i}=1/\mathcal{L}_i$ and
$\wn_{max,i}=N_{p,i}/(3\mathcal{L}_i)$, respectively, where the index
$i=\{1,2,3\}$ denotes the different directions.

In all simulations considered in this study, energy is injected through the
external force $\v{f}$ on right-hand-side of \cref{eq:ns}. The forcing scheme is
designed as proposed in \citet{Alvelius1999}; the force spectrum $F(\wn)$
is Gaussian with standard deviation $c=0.5$ and is centered around the forcing
wavenumber $\wnf$:
\begin{equation}
  F(\wn) = A \exp(-(\wn-\wnf)^2/c).
  \label{eq:forcespectrum}
\end{equation}
In \cref{eq:forcespectrum}, the prefactor $A$, which controls the amplitude of
$F(\wn)$ can be determined \textit{a priori} to the simulation and allow us to
fix the power input $\dissipI$. This is only possible because this forcing
scheme ensures that the force-velocity correlation is at all time instants zero.
As a consequence, the injected power is an exclusive product of the force-force
correlation, which is directly related to $F(\wn)$ \citep{Alvelius1999}.

\subsection{Description of the Simulations and Physical Parameters}

To describe the considered physical problem, we are free to choose $6$ control
parameters. These form the set $\{ \wnf, \dissipI, \nu, \dpar,
\dpp,\Omega\}$, which involves two physical units. Thus, a total of $4$
non-dimensional numbers is sufficient to describe the numerical experiment. The
governing non-dimensional numbers can be built by combination of the free
control parameters. For instance, using $\wnf$ and $\dissipI$ and assuming that
the constant of proportionality is $1$, we can construct the velocity scale $\uf
= \dissipI^{1/3}\wnf^{-1/3}$ and the time scale $\tf =
\wnf^{-2/3}\dissipI^{-1/3}$. Additionally, a characteristic length scale can be
taken as $\lf=1/\wnf$. Hence, the Reynolds and the Rossby number are defined as
\begin{equation}
    \reynoldseps = \frac{\dissipI^{1/3} \wnf^{-4/3}}{\nu}
    \quad \text{and} \quad
    \rossbyeps = \frac{\wnf^{2/3} \dissipI^{1/3}}{2\Omega}.
\end{equation}
The two other governing non-dimensional numbers are formed by combining the
forcing wavenumber with the geometric dimensions of the domain to yield
$\wndpp$ and $\wndpar$. The $4$ non-dimensional numbers,
$\{\reynoldseps,\rossbyeps,\wndpar,\wndpp\}$, whose definitions have been
borrowed from \citet{Seshasayanan2018}, form the parameter space henceforth used
to characterize the simulations performed in this study. Note, however, that
this set of non-dimensional parameters is not unique. For instance, one may
combine $\reynoldseps$ and $\rossbyeps$ to form the micro-scale Rossby number
$\rossbylmbda=\reynoldseps^{1/2}\rossbyeps$, which represents the ratio of
rotation and Kolmogorov time scales, or express the geometric dimensions in
terms of the domain aspect ratio $A_r = \dpar / \dpp$.

\begin{table}
\begin{center}
\def~{\hphantom{0}}
\addtolength{\tabcolsep}{.15cm}
\begin{tabular}{cccccccc}
$\wnf\mathcal{L}_\perp$  & $\wnf\mathcal{L}_\mypara$ &
$(2\pi\dpp)/\lpp^{\,\text{iso}}$ & $(2\pi\dpar)/\lpar^{\,\text{iso}}$ &
$\tf/\teddy$ & $\reynoldseps$ & $\reynoldslmbda$ & $N_p$\\[3pt]
8 & 64 & $39.9$ & $342.5$ & $2.36$ & $55.05$ &
$68$ &$768^2\times6144$\\[5pt]
\end{tabular}
\addtolength{\tabcolsep}{-.15cm}
 \caption{Numerical and physical parameters of the initial homogeneous isotropic
turbulent flow field used for the runs with rotation.}
\label{tb:initcond}
\end{center}
\end{table}

Another important parameter is the Zeman wavenumber
$\kzeman=(\Omega^3/\dissipI)^{1/2}$, which indicates the wavenumber range for
which rotational effects are relevant \citep{Zeman1994,Delache2014}. The Zeman
wavenumber is also automatically set by fixing the aforementioned parameters as
$\rossbyeps=(\wnf/\kzeman)^{2/3}/2$.

\textit{A posteriori}, we can compute the usual physical parameters that
describe the flow field. The box-averaged kinetic energy $\ken$ is given by
$\boxmean{\ui\ui}/2$, where the operator $\boxmean{\,\cdot\,}$ denotes volume
averages, and the viscous dissipation rate is
$\dissipnu=2\nu\boxmean{S_{ij}S_{ij}}$, where
$S_{ij}=(\partial{u_{i,j}}+\partial{u_{j,i}})/2$ is the strain-rate tensor. From
$\ken$, we define the r.m.s. velocity $\urms=\sqrt{2K/3}$, which is used to
define the large-eddy turnover time $\teddy=\urms^2/\dissipI$. The Taylor
micro-scale is defined as in \citet{pope_turbulent_flows}, i.e,
$\lambda=(15\nu\urms^2/\dissipnu)^{1/2}$. The Taylor micro-scale Reynolds number
is computed as $\reynoldslmbda=\urms\lambda/\nu$, and the Kolmogorov
length scale is $\lkolmo=(\nu^3/\dissipI)^{1/4}$.

Last, we define the integral length scales along the directions normal and
parallel to the axis of rotation. These are constructed from the two-point
velocity correlation:
\begin{equation}
    R(\v{r}) = \frac{\mean{\ui(\v{x}) \ui(\v{x}+\v{r})}_\mathcal{L}}
{\mean{\ui(\v{x}) \ui(\v{x})}_\mathcal{L}},
\label{eq:twopoint}
\end{equation}
where $\v{r}=r_i\uv{e}_i$ is an arbitrary position vector. We integrate
\cref{eq:twopoint} with $\v{r}= r \uv{e}_r$, as in spherical coordinates, or
with $\v{r}= \rpp \uv{e}_\perp$ and $\v{r}= \rpar\uv{e}_\mypara$, as in
cylindrical coordinates, to obtain the integral length scales along the
respective directions:
\begin{equation}
  \ell = \int_0^{\pi \mathcal{L}_{min}} R(r) \diff{r},
  \quad
  \lpp = \int_0^{\pi\dpp} R(\rpp) \diff{\rpp},
  \quad \text{and} \quad
  \lpar = \int_0^{\pi \dpar} R(\rpar) \diff{\rpar}.
  \label{eq:defintscales}
\end{equation}
In \cref{eq:defintscales}, $\mathcal{L}_{min}$ is taken as $\min(\dpar,\dpp)$ in
the limit of the integral that defines $\ell$. To represent quantities from the
initial and isotropic flow field we use the superscript ``${\text{iso}}$'', like
in $\ell^{\,\text{iso}}$.

\begin{figure}
\centering
\captionsetup[subfigure]{labelformat=empty}
\subfloat[\label{fig:pp_par_correlation}]{}
\subfloat[\label{fig:1d_energy_spectra}]{}
\begin{tikzpicture}
\coordinate (fig1pos) at (0,0);
\coordinate (fig1xlabel) at ([yshift=-2.7cm]fig1pos);
\coordinate (fig1ylabel) at ([xshift=-3.2cm]fig1pos);
\coordinate (caption)    at ([xshift=-3.0cm,yshift=2.25cm]fig1pos);
\node[inner sep=0pt] (figleft) at (fig1pos){
      \includegraphics[width=.48\linewidth]
      {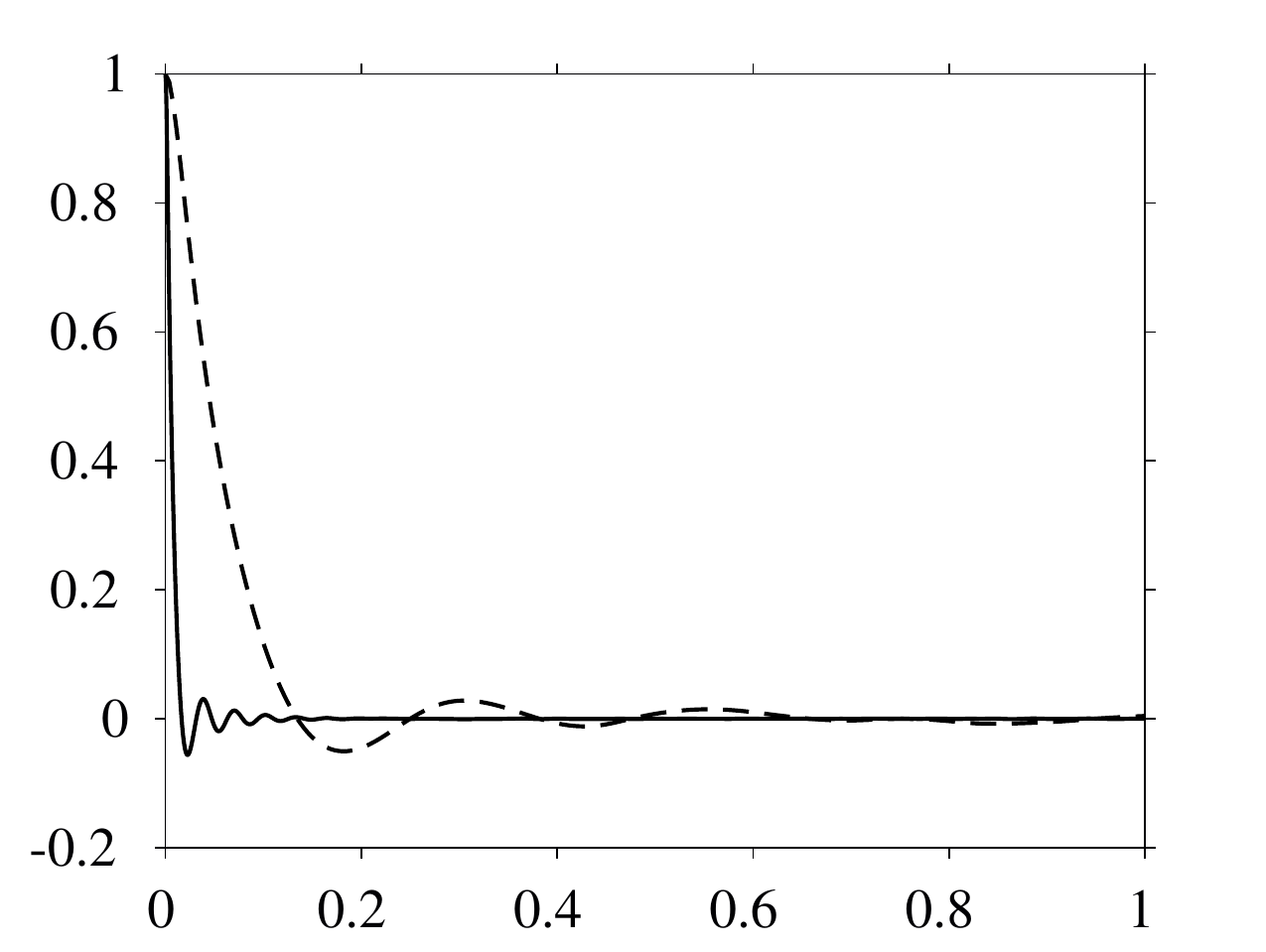}
  };
\node            (xfig1) at (fig1xlabel) {$r_\perp/(\pi \mathcal{L}_\perp)
                       \quad\text{;}\quad r_\mypara/(\pi \mathcal{L}_\mypara)$};
\node[rotate=90] (yfig1) at (fig1ylabel) {
                                         $R(\rpp) \quad\text{;}\quad R(\rpar)$};
\node            (figa)  at (caption) {(a)};
\coordinate (fig2pos) at (6.4,0);
\coordinate (fig2xlabel) at ([yshift=-2.7cm]fig2pos);
\coordinate (fig2ylabel) at ([xshift=-3.3cm]fig2pos);
\coordinate (caption)    at ([xshift=-3.2cm,yshift=2.25cm]fig2pos);
\node[inner sep=0pt] (figleft) at (fig2pos){
      \includegraphics[width=.48\linewidth]
      {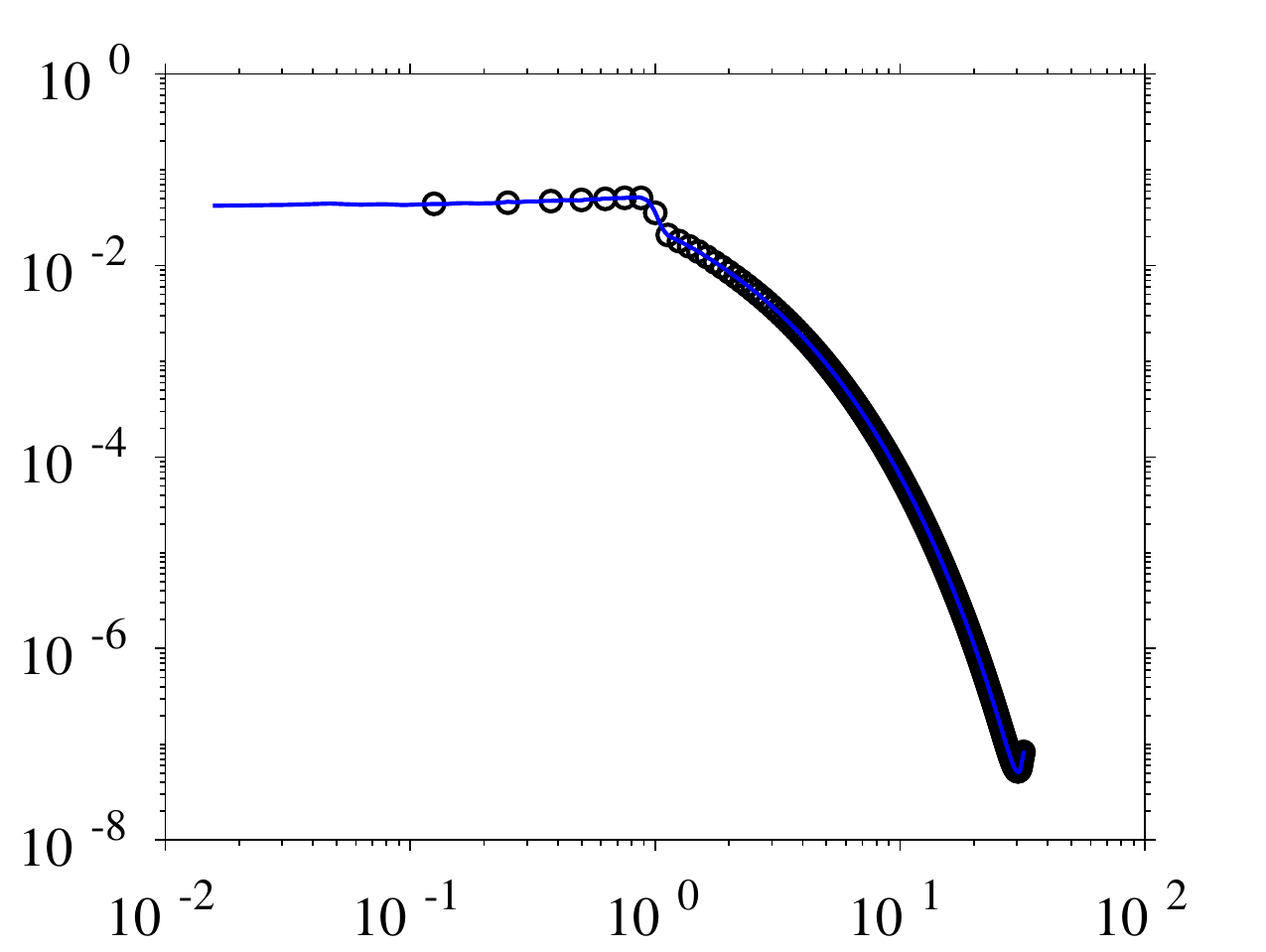}
  };
\node            (xfig2) at (fig2xlabel) {$\wn_\beta/\wnf$};
\node[rotate=90] (yfig2) at (fig2ylabel) {$E_{\alpha\alpha}(\wn_\beta)$};
\node            (figb)  at (caption) {(b)};
\end{tikzpicture}
 \caption{Two-point velocity correlations showing the ratio domain size to
characteristic size of the flow structures in the different directions, and
one-dimensional energy spectra showing that the initial conditions are indeed
isotropic. (a) Normal $R(\rpp)$ \mbox{(\protect\blackdashed)} and parallel
$R(\rpar)$ \mbox{(\protect\blackline)} velocity two-point correlations. (b)
Perpendicular and parallel one-dimensional energy spectra: $\alpha=1$;
$\beta=3$; $E_{11}(\wn_3)$ \mbox{(\protect\blueline)} and  $\alpha=3$; $\beta=1$;
$E_{33}(\wn_1)$
\mbox{(\protect\oblack)}}
\label{fig:corr_1d_spectra}
\end{figure}
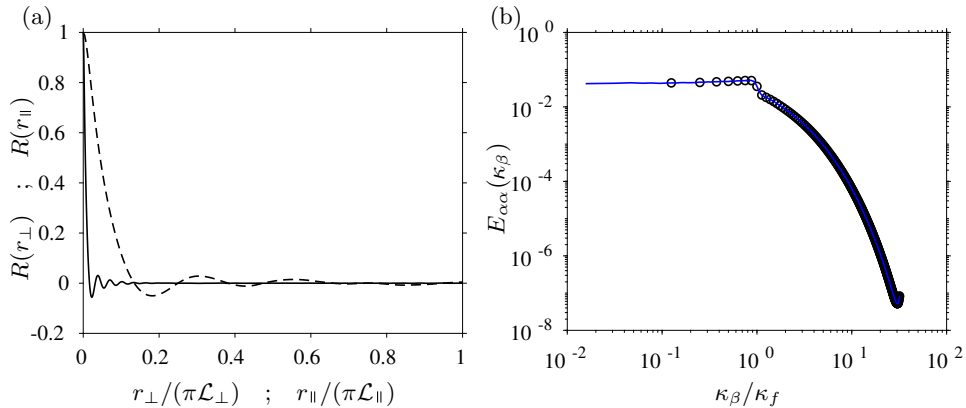

\begin{table}
\begin{center}
\def~{\hphantom{0}}
\addtolength{\tabcolsep}{.3cm}

\begin{tabular}{ll}
& \begin{tabular}{c@{\hskip 0.4cm}c@{\hskip 0.4cm}cc@{\hskip 1cm}cc}
Run & Colormap & $\rossbyeps$ &$\rossbylmbda$ &$\kzeman\eta$ &
$\kzeman/\wnf$
\end{tabular}
\\[5pt]
R1 & $\left\{
    \begin{tabular}
    {@{}llllll}
\texttt{run01} & \textcolor{run01}\straightline & $1.54$ & $11.42$ & $0.009$
    & $0.19$
\tabularnewline
\texttt{run02} & \textcolor{run02}\straightline & $1.25$ & $9.28$ & $0.013$
    & $0.25$
\tabularnewline
\texttt{run03} & \textcolor{run03}\straightline & $1.00$ & $7.42$ & $0.017$
    & $0.35$
\tabularnewline
\texttt{run04} & \textcolor{run04}\straightline & $0.87$ & $6.45$ & $0.022$
    & $0.44$
\tabularnewline
\texttt{run05} & \textcolor{run05}\straightline & $0.77$ & $5.71$ & $0.026$
    & $0.52$
\tabularnewline
\texttt{run06} & \textcolor{run06}\straightline & $0.69$ & $5.12$ & $0.031$
    & $0.62$
\tabularnewline
\texttt{run07} & \textcolor{run07}\straightline & $0.63$ & $4.64$ & $0.035$
    & $0.72$
\tabularnewline
\texttt{run08} & \textcolor{run08}\straightline & $0.56$ & $4.12$ & $0.042$
    & $0.85$
\tabularnewline
\texttt{run09} & \textcolor{run09}\straightline & $0.47$ & $3.45$ & $0.055$
    & $1.11$
\tabularnewline
\texttt{run10} & \textcolor{run10}\straightline & $0.39$ & $2.91$ & $0.071$
    & $1.44$
    \end{tabular}\right.\kern-\nulldelimiterspace$
\tabularnewline\tabularnewline
R2 & $\left\{
    \begin{tabular}{@{}llllll}
\texttt{run11} & \textcolor{run11}\straightline & $0.31$ & $2.32\phantom{0}$
    & $0.100$ & $2.02$
\tabularnewline
\texttt{run12} & \textcolor{run12}\straightline & $0.27$ & $2.01$ & $0.124$
    & $2.52$
\tabularnewline
\texttt{run13} & \textcolor{run13}\straightline & $0.24$ & $1.79$ & $0.148$
    & $2.99$
\tabularnewline
\texttt{run14} & \textcolor{run14}\straightline & $0.22$ & $1.60$ & $0.175$
    & $3.55$
\tabularnewline
\texttt{run15} & \textcolor{run15}\straightline & $0.19$ & $1.40$ & $0.213$
    & $4.31$
\tabularnewline
\texttt{run16} & \textcolor{run16}\straightline & $0.16$ & $1.20$ & $0.270$
    & $5.46$
\tabularnewline
\texttt{run17} & \textcolor{run17}\straightline & $0.14$ & $1.00$ & $0.352$
    & $7.12$
\tabularnewline
\texttt{run18} & \textcolor{run18}\straightline & $0.11$ & $0.80$ & $0.492$
    & $9.95$
\tabularnewline
\texttt{run19} & \textcolor{run19}\straightline & $0.09$ & $0.70$ & $0.599$
    & $12.12$
\tabularnewline
\texttt{run20} & \textcolor{run20}\straightline & $0.08$ & $0.60$ & $0.759$
    & $15.34$
\tabularnewline
\texttt{run21} & \textcolor{run21}\straightline & $0.06$ & $0.47$ & $1.088$
    & $21.99$
    \end{tabular}\right.\kern-\nulldelimiterspace$
\tabularnewline\tabularnewline
\end{tabular}
\addtolength{\tabcolsep}{-.3cm}
 \caption{Numerical and physical parameters for the DNS of homogeneous
rotation turbulence at distinct rates of rotation. The runs that exhibit
similar dynamics are collected together in groups, namely $R1$ and $R2$.}
\label{tb:perfosimul}
\end{center}
\end{table}

\begin{figure}
\centering
\captionsetup[subfigure]{labelformat=empty}
\subfloat[\label{fig:visualization_iso}]{}
\subfloat[\label{fig:visualization_t1}]{}
\subfloat[\label{fig:visualization_t2}]{}
\begin{tikzpicture}
\coordinate (fig1pos)   at (0,0);
\coordinate (fig2pos)   at (0,0);
\coordinate (fig2pos)   at (0,0);
\coordinate (captiona)  at (-4.8,5.8);
\coordinate (captionb)  at (0,5.8);
\coordinate (captionc)  at (4.5,5.8);
\coordinate (origin)    at (-0.25,-9);
\coordinate (originomg) at (0.4 ,-8.5);
\coordinate (ox1)       at ([xshift=1.19cm,yshift=0.12cm]origin);
\coordinate (ox2)       at ([xshift=-1.0cm,yshift=0.4cm]origin);
\coordinate (ox3)       at ([xshift=0cm,yshift=1.2cm]origin);
\coordinate (oxmg)      at ([xshift=0.65cm,yshift=1.2cm]origin);
\node[inner sep=0pt] (figa) at (fig1pos){
      \includegraphics[trim=2cm 10cm 0 0,clip,width=.31\linewidth]
      {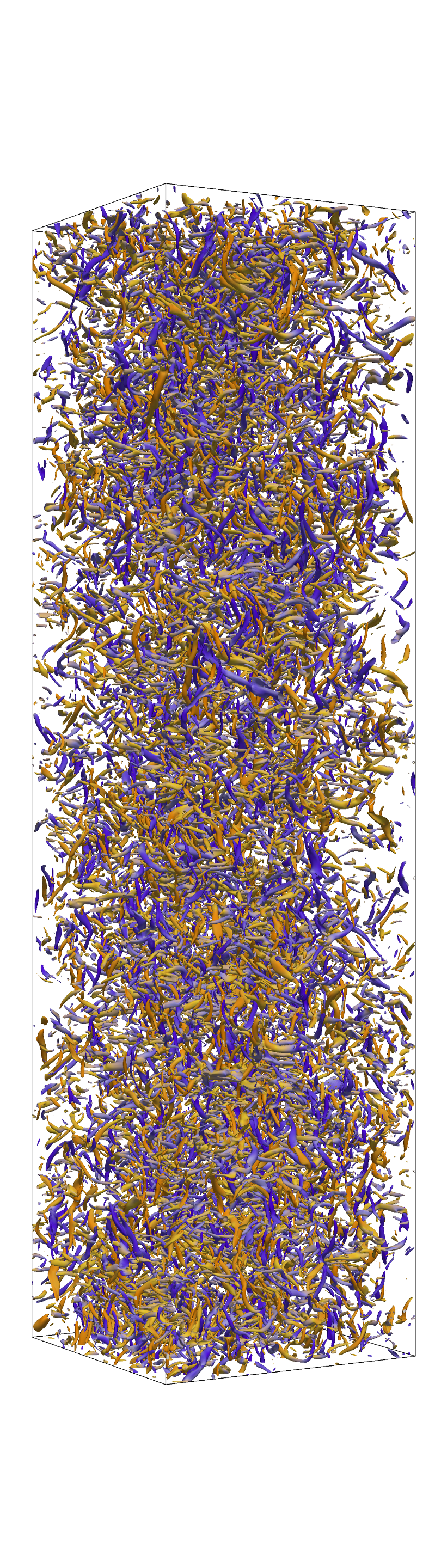}
      \quad
      \includegraphics[trim=2cm 10cm 0 0,clip,width=.31\linewidth]
      {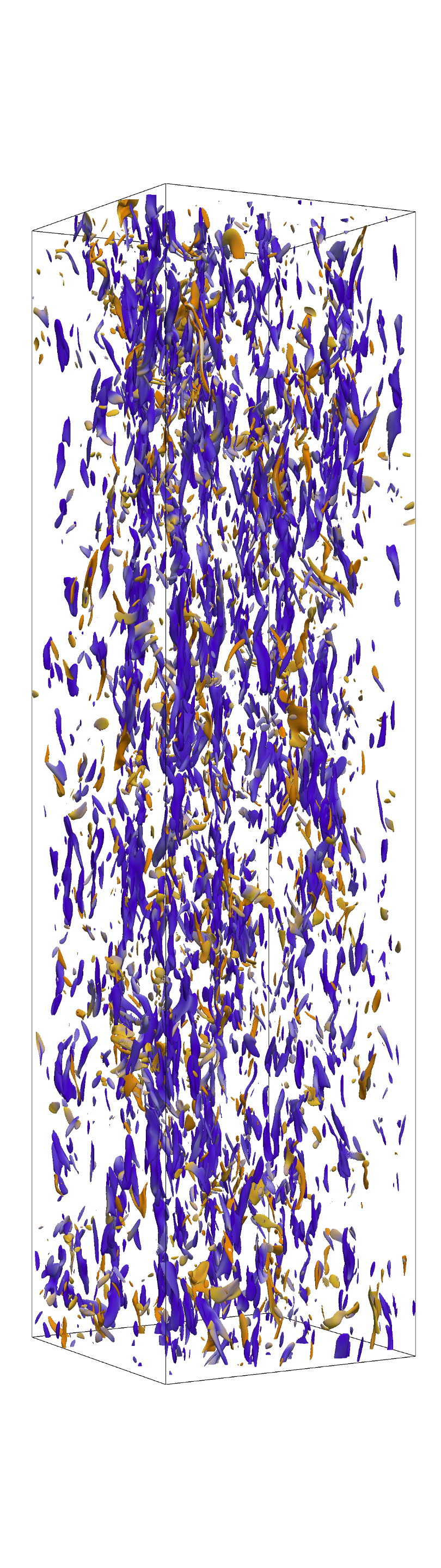}
      \quad
      \includegraphics[trim=2cm 10cm 0 3cm,clip,width=.31\linewidth]
      {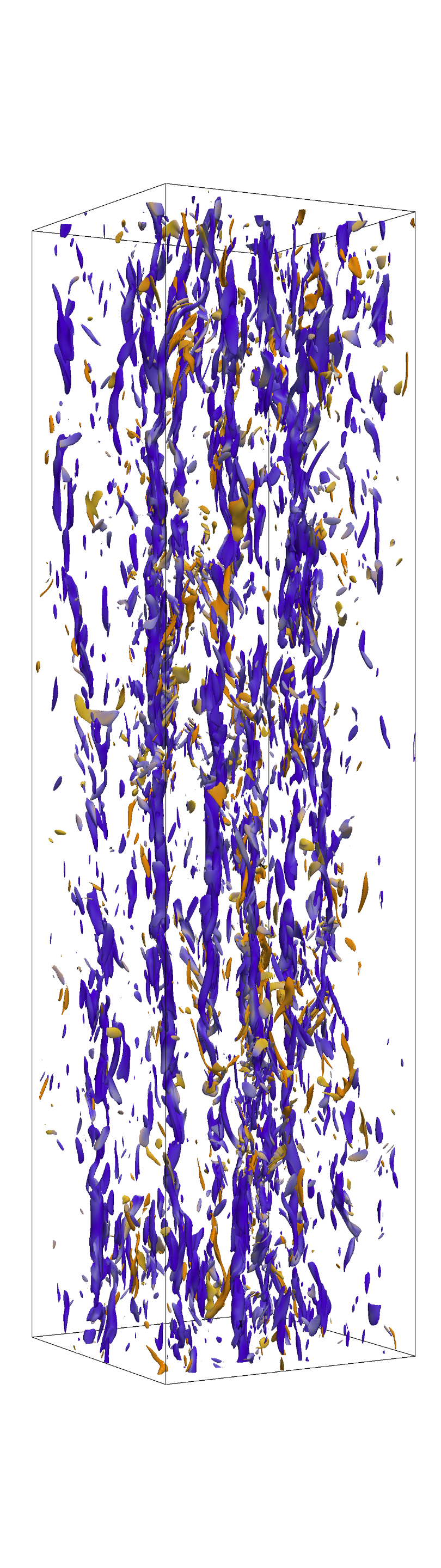}
      };
\node at (captiona) {(a)};
\node at (captionb) {(b)};
\node at (captionc) {(c)};
\draw[-latex,line width=.3mm] (origin) -- (ox1);
\draw[-latex,line width=.3mm] (origin) -- (ox2);
\draw[-latex,line width=.3mm] (origin) -- (ox3);
\draw[-latex,line width=.3mm] (originomg) -- (oxmg);
\node at ([xshift=0.2cm,yshift=0.2cm]ox1) {$x_1$};
\node at ([xshift=0.2cm,yshift=0.2cm]ox2) {$x_2$};
\node at ([xshift=0.2cm,yshift=0.1cm]ox3) {$x_3$};
\node at ([xshift=0.3cm,yshift=-0.1cm]oxmg) {$\gv{\Omega}$};
\end{tikzpicture}
 \caption{Flow field visualization of a sub-set of the computational domain
($1/16$ of the entire computational domain), showing half of the horizontal
domain extension and $1/4$ of the vertical domain size:
$[0,\pi]\times[0,\pi]\times[0,4\pi]$. Iso-contours of the Q-criterion
\citep{Hunt1988} colored by the normalized projection of the vorticity vector
along the axis of rotation, i.e., $\vort\cdot\v{e}_\mypara/\norm{\vort}$. Blue
colors indicates structures that rotate in the same sense as $\gv{\Omega}$
(counter-clockwise), whereas orange colors indicate the opposite sense of
rotation (clockwise). (a) Isotropic initial condition; (b) and (c) correspond to
the run with $\rossbyeps=0.06$ at later time instants after the onset of
rotation, i.e., $t=10.5\,\tf$ and $t=20\,\tf$, respectively.}
\label{fig:visualization}
\end{figure}
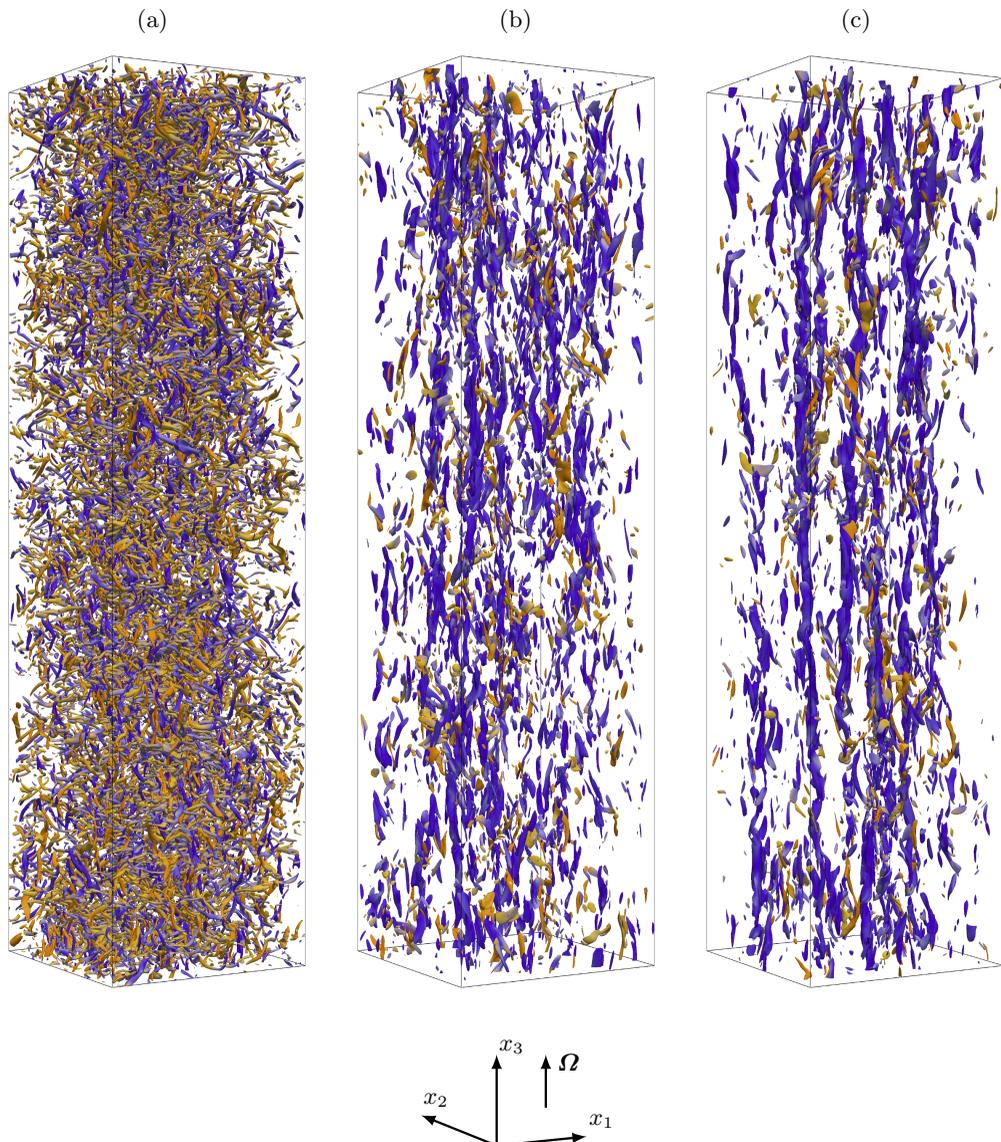

The initial conditions for the simulations with rotation are produced by
injecting energy at constant rate $\dissipI$ to a fluid that is initially at
rest. The energy, which is injected at wavenumber $\wnf=8$, is progressively
distributed over a wider range of wavenumbers by action of the triple velocity
correlations. When the energy cascade is built-up, the box-averaged kinetic
energy $\ken$ stops growing and a steady-state is reached.  The numerical
resolution guarantees that at all times $\wn_{max}\eta\ge1.5$, which is
sufficient to resolve all scales of motion. The initial transient lasts for
$20\,\tf$ or, equivalently, $8.45\,\teddy$, and afterwards, statistics are
collected for another $54\,\tf$ ($22.84\,\teddy$). For the fully developed
field, we find that $\reynoldslmbda\approx68$, and that the relation
$\liso=\lpariso=\lppiso$ holds up to $2$ decimal places. The latter suggests
that the flow field is in fact isotropic.

Other statistics of the steady-state match closely with typical values found in
DNS of homogeneous isotropic turbulence. For instance, the skewness and
flatness of the longitudinal velocity derivative $\pdinline{u_1}{x_1}$ are
$-0.51$ and $4.8$, respectively, in agreement with \citet{Tang2018} and
\citet{VanAtta1980}. The energy dissipation rate $\dissipnu$ at the steady-state
is well approximated by $\dissipnu=\cepsiso(\urmsiso)^3/\liso$, where
$\cepsiso\approx0.35$ is the constant of proportionality. Note, however, that
the value of this constant depends on how the two-point correlation in
\cref{eq:twopoint} is normalized. If we normalize it with $2\urms$, like in
\citet{Kaneda2003}, instead of $2\ken$, like in \cref{eq:twopoint}, a factor of
$3/2$ must be accounted for to yield $\cepsiso\approx0.5$ in agreement with
literature; see \citet{Ishihara2009b} for a compilation of other numerical
results.

In this study, the goal is not to achieve the highest possible Reynolds number
for a given numerical resolution. Instead, we focus on maximizing the time for
which large scale eddies with typical size $\liso$ can evolve unbounded, while
still resolving all scales of motion. Therefore, apart from forcing at scales
smaller than usual, we consider an elongated domain with $A_r=8$. As a result,
the isotropic fields to which background rotation can be imposed to are in the
vertical direction about $340$ times larger than $\liso$ and, in the normal
direction, $2\pi\dpp/\liso\approx40$. In \cref{fig:corr_1d_spectra}, we show
evidence of these aspects. \Cref{fig:pp_par_correlation} confirms through the
two-point velocity correlation along the normal and the parallel directions that
the ratio domain size to flow structures is indeed significantly larger in the
vertical direction. Alongside, \cref{fig:1d_energy_spectra} verifies that the
velocity fields are isotropic, as the curves for the one-dimensional energy
spectra along the normal and perpendicular directions overlap.

These features are also clearly visible in the flow field visualization; see
\cref{fig:visualization}, where we show a sub-set of the computational domain
with the flow structures visualized by the Q-Criterion of \citet{Hunt1988} and
colored by the normalized projection of the vorticity vector on the axis of
rotation, i.e., $\vort\cdot\v{e}_\mypara/\norm{\vort}$. Reinforcing the
aforementioned results, we observe two main points in the isotropic field that
is used as initial condition for the runs with rotation
(\cref{fig:visualization_iso}). First, the flow structures do not display any
preferential sense of rotation, which is confirmed by the uniform distribution
of the colors. Second, they are also isotropically arranged and therefore not
aligned along any preferential direction. For a summary of the numerical and
physical parameters of the initial conditions, please refer to
\cref{tb:initcond}.

The runs with rotation are constructed by imposing $21$ different background
rotation rates to the isotropic flow field shown in
\cref{fig:visualization_iso}; see \cref{tb:perfosimul} for the relevant
numerical and physical parameters. The result is a set of simulations that
covers a broad range of the $\rossbyeps$ parameter space, i.e.,
$0.06>\rossbyeps>1.54$. The Zeman wavenumber in terms of the Kolmogorov
length scale, $\kzeman\eta$, for instance, varies from $0.1$ for
$\rossbyeps=1.54$ (weakest rotation case) to $1.1$ for $\rossbyeps=0.06$
(strongest rotation case). As the numerical resolution provides
$\wn_{max}\eta=1.5$ for the fully developed isotropic reference initial field,
for $\rossbyeps=0.06$, almost all scales of motion are influenced by the
system's rotation.

The effects of rotation on the flow structures are readily seen in the
flow-field visualization in \cref{fig:visualization}. Together with the initial
conditions, we see in \cref{fig:visualization_t1,fig:visualization_t2} two
snapshots for the run with $\rossbyeps=0.06$ at times subsequent to the onset of
rotation ($t=10.5\,\tf$ and $t=20\,\tf$). In agreement with
common knowledge, we confirm that the rotation destroys the small
structures and modulates the flow field such that columns elongated in the
direction of rotation emerge. At the later instant of time
(\cref{fig:visualization_t2}), structures in blue predominate, indicating that
flow structures that rotate in the same sense as the imposed background rotation
are more likely to be found. This is commonly referred to as the cyclone
asymmetry and has been observed in the experiments of \citet{VanBokhoven2009}
and in the computations of \citet{bartello:1994}.

\begin{figure}
\centering
\captionsetup[subfigure]{labelformat=empty}
\subfloat[\label{fig:pp_lscale_weak}]{}
\subfloat[\label{fig:pp_lscale_strong}]{}\subfloat[\label{fig:par_lscale_weak}]{}
\subfloat[\label{fig:par_lscale_strong}]{}
\begin{tikzpicture}
\coordinate (fig1pos) at (0,0);
\coordinate (fig1xlabel) at ([yshift=-2.7cm]fig1pos);
\coordinate (fig1ylabel) at ([xshift=-3.4cm]fig1pos);
\coordinate (caption)    at ([xshift=-3.2cm,yshift=2.25cm]fig1pos);
\node[inner sep=0pt] (figleft) at (fig1pos){
      \includegraphics[width=.48\linewidth]
      {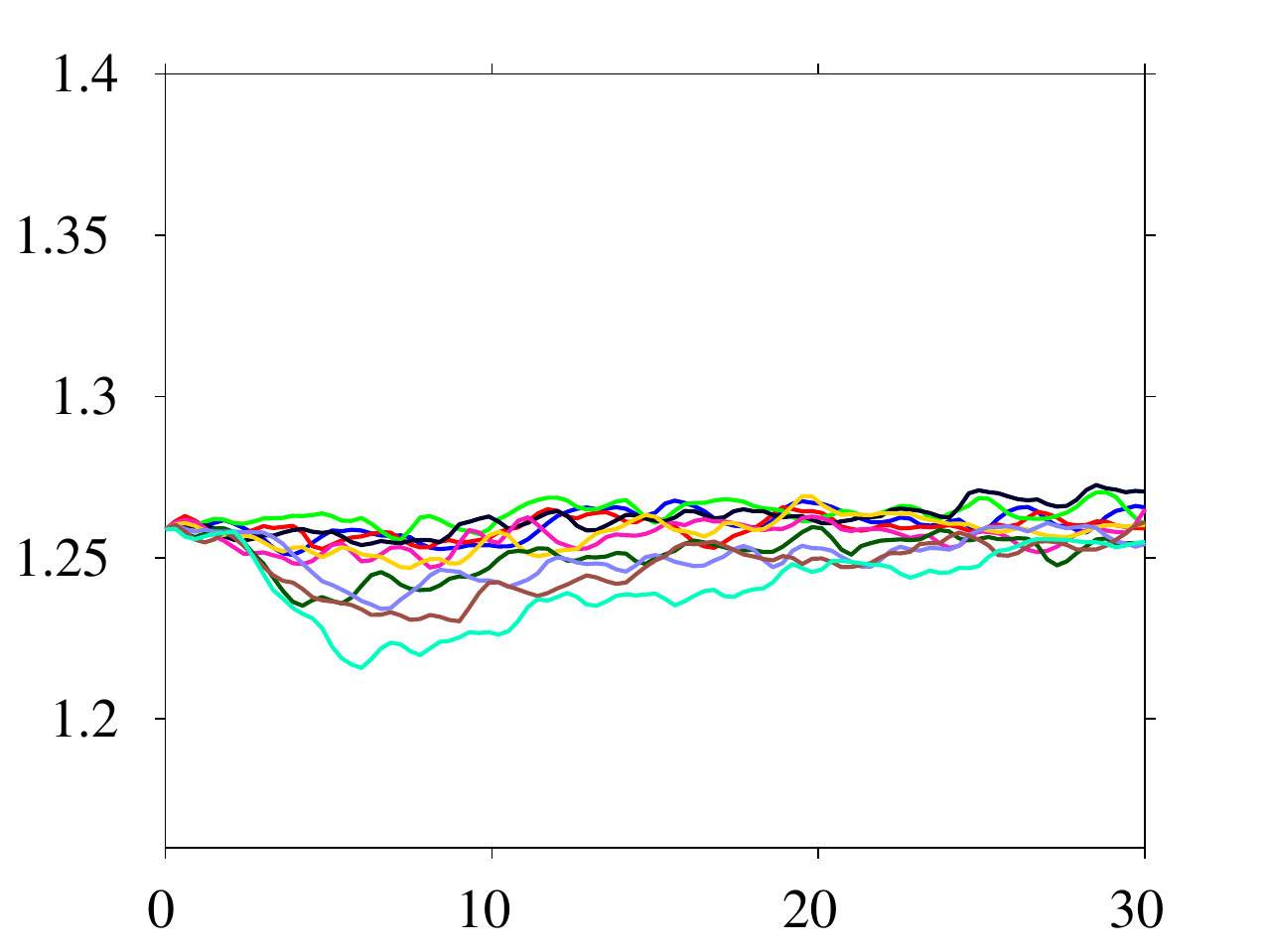}
  };
\node            (xfig1) at (fig1xlabel) {$t/\tf$};
\node[rotate=90] (yfig1) at (fig1ylabel) {$\wnlpp$};
\node            (figa)  at (caption)    {(a)};
\coordinate (fig2pos) at (6.4,0);
\coordinate (fig2xlabel) at ([yshift=-2.7cm]fig2pos);
\coordinate (fig2ylabel) at ([xshift=-3.2cm]fig2pos);
\coordinate (caption)    at ([xshift=-3.2cm,yshift=2.25cm]fig2pos);
\node[inner sep=0pt] (figleft) at (fig2pos){
      \includegraphics[width=.48\linewidth]
      {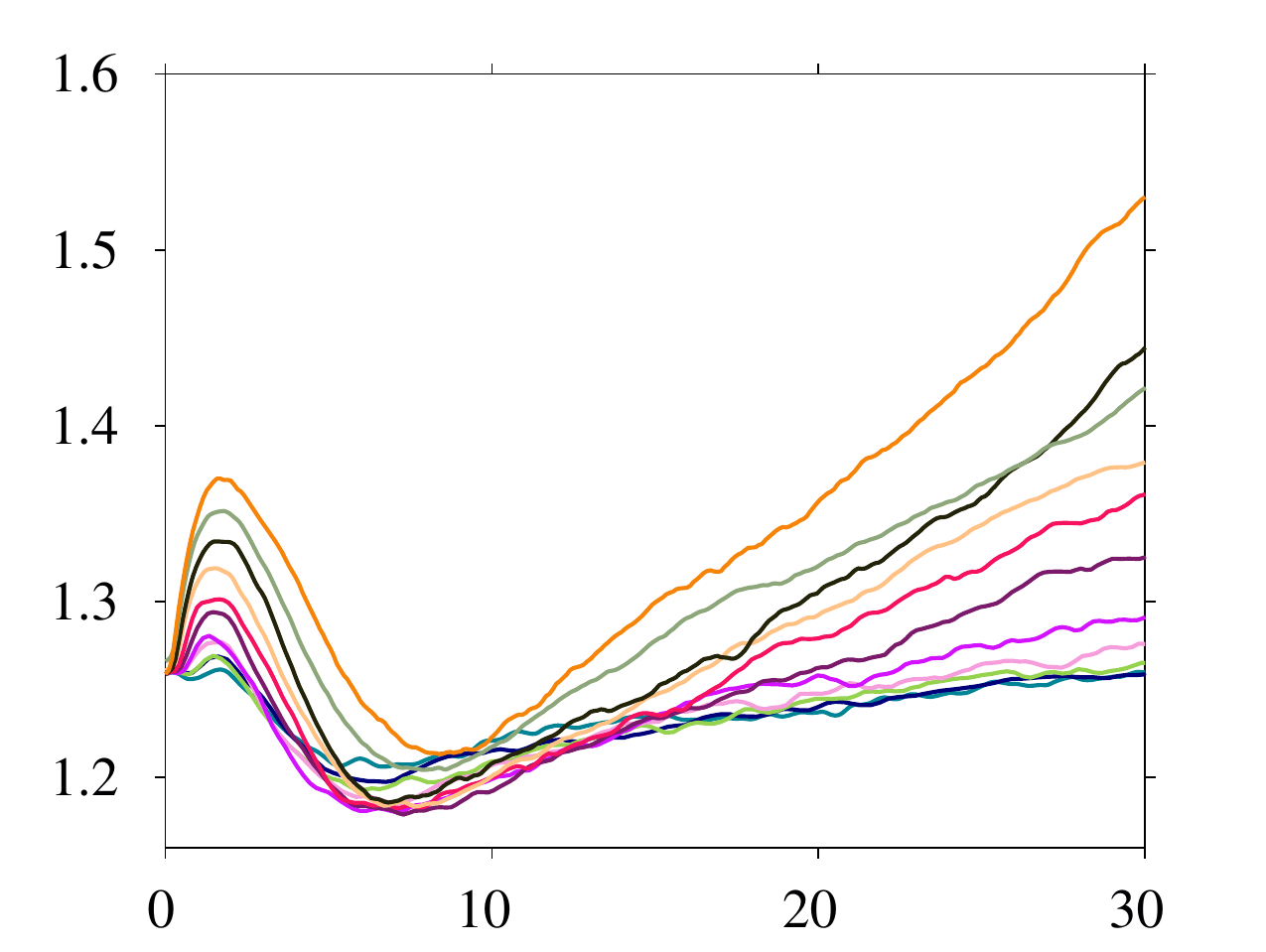}
  };
\node            (xfig1) at (fig2xlabel) {$t/\tf$};
\node[rotate=90] (yfig1) at (fig2ylabel) {$\wnlpp$};
\node            (figb)  at (caption)    {(b)};
\coordinate (fig3pos) at (0,-5.5);
\coordinate (fig3xlabel) at ([yshift=-2.7cm]fig3pos);
\coordinate (fig3ylabel) at ([xshift=-3.4cm]fig3pos);
\coordinate (caption)    at ([xshift=-3.2cm,yshift=2.25cm]fig3pos);
\node[inner sep=0pt] (figleft) at (fig3pos){
      \includegraphics[width=.48\linewidth]
      {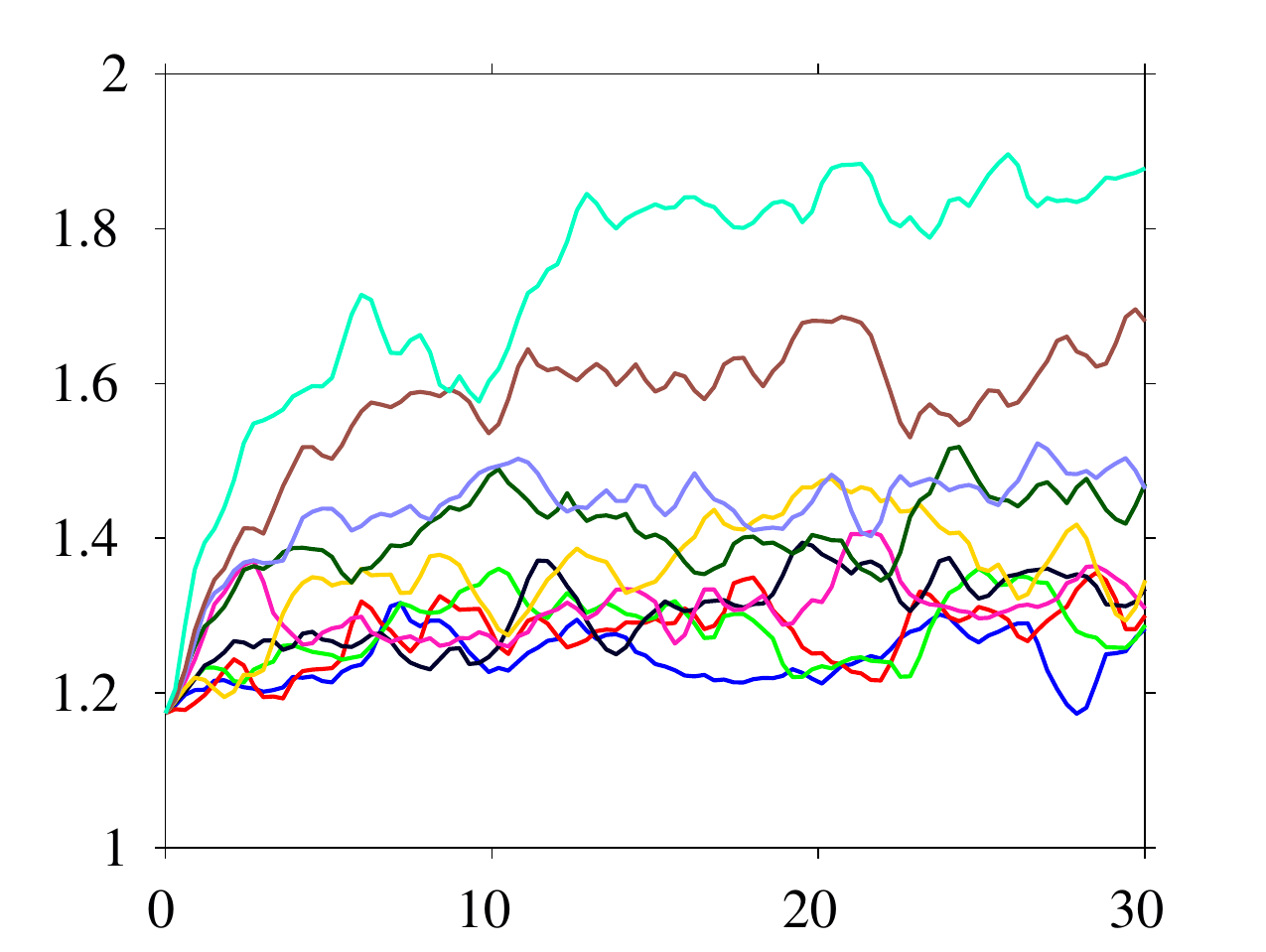}
  };
\node            (xfig3) at (fig3xlabel) {$t/\tf$};
\node[rotate=90] (yfig3) at (fig3ylabel) {$\wnlpar$};
\node            (figc)  at (caption)    {(c)};
\coordinate (fig4pos) at (6.4,-5.5);
\coordinate (fig4xlabel) at ([yshift=-2.7cm]fig4pos);
\coordinate (fig4ylabel) at ([xshift=-3.2cm]fig4pos);
\coordinate (caption)    at ([xshift=-3.2cm,yshift=2.25cm]fig4pos);
\node[inner sep=0pt] (figleft) at (fig4pos){
      \includegraphics[width=.48\linewidth]
      {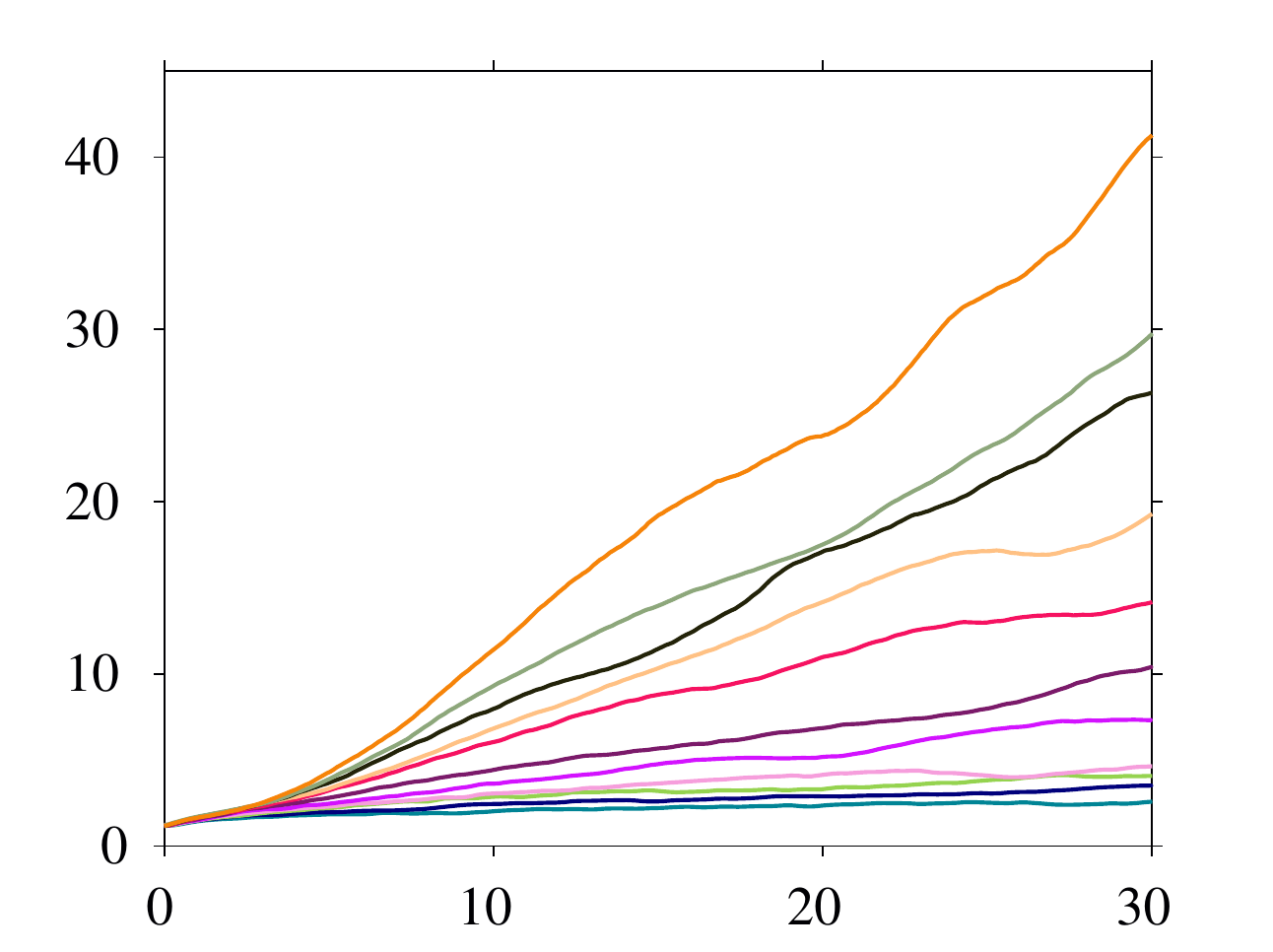}
  };
\node            (xfig4) at (fig4xlabel) {$t/\tf$};
\node[rotate=90] (yfig4) at (fig4ylabel) {$\wnlpar$};
\node            (figb)  at (caption)    {(d)};
\end{tikzpicture}
 \caption{Time evolution of the parallel and transversal integral length scales
$\lpar$ and $\lpp$, for $1.56\ge\rossbyeps\ge0.39$ (left panels) and
$0.31 \ge\rossbyeps\ge 0.06$ (right panels)}
\label{fig:int_lscales}
\end{figure}

\section{The Growth Rate of Columnar Eddies}
\label{sec:growthlpar}

Now, we present results and discuss the influence of different rotation rates on
the growth of the columnar eddies. For the quantitative analysis, we use
integral length scales, which on one hand can be used to quantify the typical
eddy size that contributes the most to the total kinetic energy, and on the
other hand also serves as an indicator of anisotropy. Due to the background
rotation, the dynamics of the flow in the parallel and transversal direction are
essentially different, which is reflected in the temporal evolution of $\lpar$
and $\lpp$ \citep{Rogallo1985}. As it will be seen, the appearance of the
columnar eddies in \cref{fig:visualization_t1,fig:visualization_t2} is strongly
reflected in the growth of the integral length scale along the axis of rotation.

We obtain the time evolution of $\lpar$ and $\lpp$ by evaluating
\cref{eq:defintscales} on a series of instantaneous velocity fields throughout
the simulation time, see \cref{fig:int_lscales}. We choose to split the actual
data in two diagrams, which are displayed side-by-side. The left panels
correspond to cases for which $\rossbyeps\ge0.39$ (group $R1$ in
\cref{tb:perfosimul}) and the right panels to $\rossbyeps\le0.31$ (group $R2$ in
\cref{tb:perfosimul}). We organized the results in this manner, so cases with
similar dynamics are shown together.

For $\rossbyeps\ge0.39$ (left panels; group $R1$), $\lpar$ and $\lpp$ remain
approximately unchanged in time and at values similar to the ones at $t=0$,
which corresponds to the initial isotropic field. Specifically for
$\rossbyeps=0.39$, the run with highest rotation rate in this group, the
departure from isotropy is marginal and $\lpar/\lpp\approx1.5$ at the final
simulation time. Differently, for $\rossbyeps\le0.31$ (right panels; group
$R2$), the disparity between $\lpar$ and $\lpp$ is clear. We observe that
$\lpar$ grows substantially in time, whereas variations in $\lpp$ are small when
compared to the latter. For instance, for $\rossbyeps=0.06$, the final value of
$\lpar$ is $26.8$ times greater that its initial value, whereas $\lpp$ only
increases by a factor of $1.21$. Additionally, we observe an intriguing behavior
in $\lpp$. It initially grows in time until a maximum is reached;
thereupon, it decreases towards a minimum, before growing again. On the other
hand, $\lpar$ increases monotonically and approximately linearly for
$t>10\,\tf$.

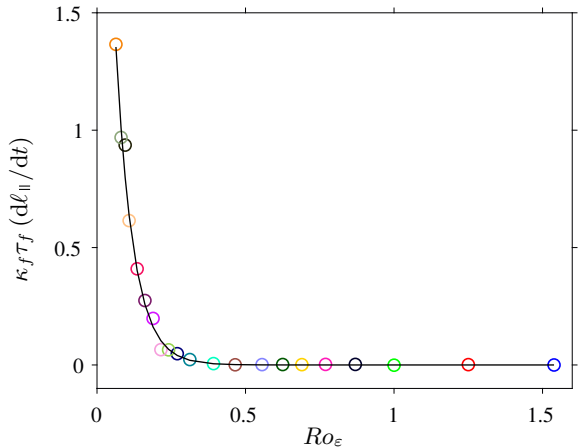
\begin{figure}
\centering
\begin{tikzpicture}
\coordinate (fig1pos) at (0,0);
\coordinate (fig1xlabel) at ([yshift=-3.0cm]fig1pos);
\coordinate (fig1ylabel) at ([xshift=-4cm]fig1pos);
\node[inner sep=0pt] (figleft) at (fig1pos){
      \includegraphics[width=.6\linewidth]
      {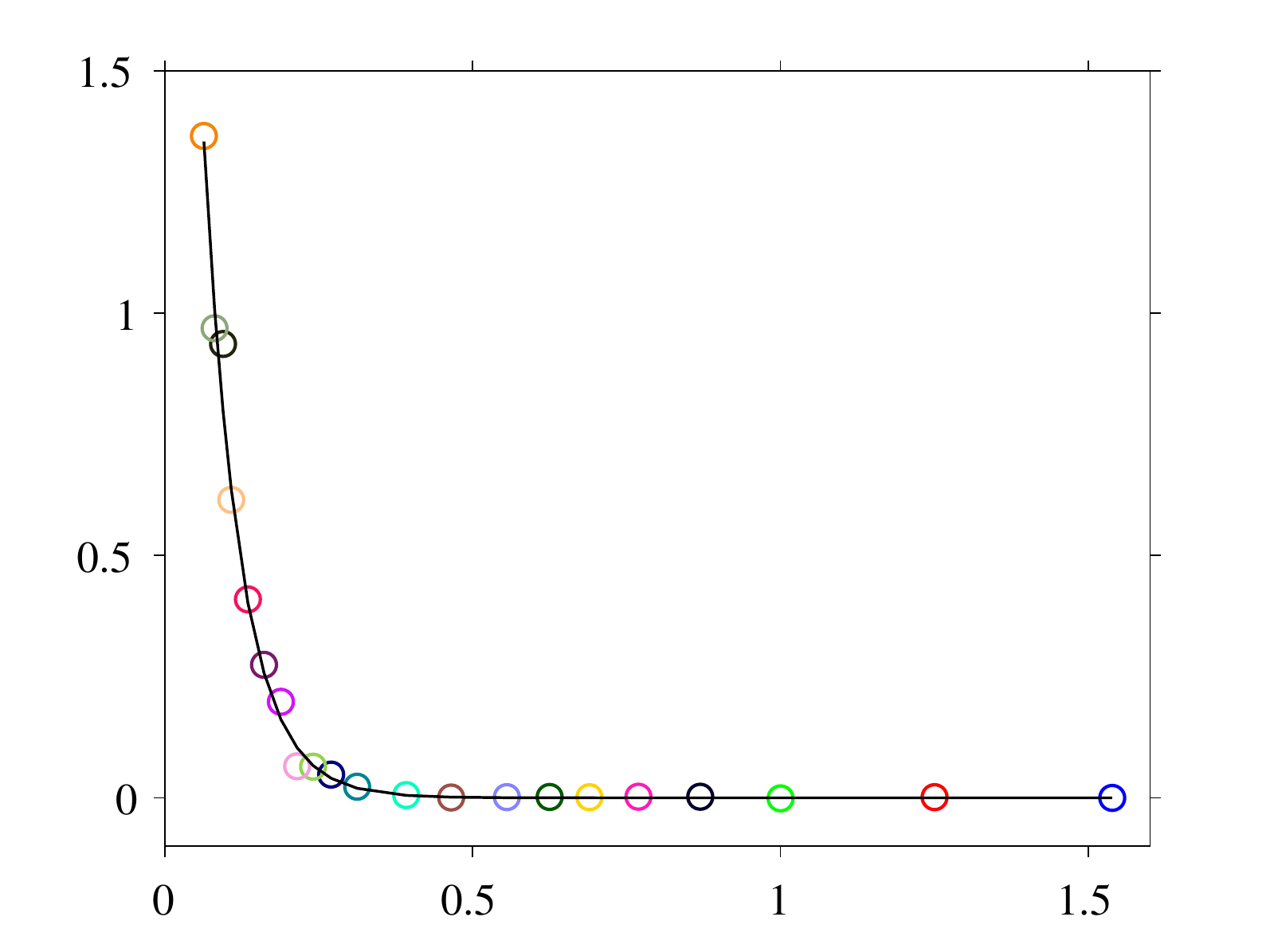}
  };
\node            (xfig1) at (fig1xlabel) {$\rossbyeps$};
\node[rotate=90] (yfig1) at (fig1ylabel){$\wnf \tf \, (\dinline{\lpar}{t})$};
\end{tikzpicture}
 \caption{Growth rate of $\lpar$ in terms of $\rossbyeps$, obtained after linear
regression in the interval $10\,\tf<t<30\,\tf$ of
\cref{fig:par_lscale_weak,fig:par_lscale_strong}. The residual of the linear fit
is maximum for $\rossbyeps=0.11$ and around $4.7\%$ of the mean value of $\lpar$
in the same interval. The blue solid line represents a exponential dependence on
$\rossbyeps$ of the type $a\exp{(b\,\rossbyeps)}$, where $a=4$ and $b=-17$
\mbox{(\protect\blueline)}.}
\label{fig:alpha_par_lscale}
\end{figure}

The growth of $\lpar$ in \cref{fig:par_lscale_strong} is in agreement with the
formation of columnar eddies observed in \cref{fig:visualization_t2}. In order
to identify the dependency between the growth rate of $\lpar$ and $\rossbyeps$,
we have fit the data for $\lpar$ in the interval $10\,\tf<t<30\,\tf$ with a
straight line. The linear fit approximates fairly well the time evolution of
$\lpar$ and the maximum residuum is found for $\rossbyeps=0.11$, where the
discrepancy is around $4.7\%$ of the mean value of $\lpar$. The slope of the
linear fit non-dimensionalized with the forcing parameters, i.e., $\gamma = \wnf
\tf (\dinline{\lpar}{t})$, is shown in \cref{fig:alpha_par_lscale} as a function
of $\rossbyeps$. For $\rossbyeps\ge0.39$, the effects of rotation are irrelevant
and $\gamma$ approaches zero, suggesting that the integral length scales remain
approximately at their initial value. On the other hand, the range
$\rossbyeps\le0.31$ is marked by a significant rise in $\gamma$. Overall, we
find that the exponential dependency $\gamma = a
\exp{(b\,\rossbyeps)}$ with $a=4$ and $b=-17$ reproduces our data-set best.

Throughout this section, we have used integral length scales based on two-point
velocity correlations to refer to vortex elongation. However, we must bear in
mind the results obtained by \citet{Yoshimatsu2011}, who showed that although a
compact spherical vortex blob subjected to pure linear dynamics grows
preferentially along the axis of rotation, $\lpar$ did not display any
substantial growth. This conundrum was attributed to the fact that integral
length scales based on the velocity field, such as $\lpar$ and $\lpp$, are fully
determined by the amplitude of the diagonal elements of the velocity spectrum
tensor. In other words, $\lpar$ and $\lpp$ lack phase information. As
alternative, it was then proposed to build integral length scales from nonlinear
quantities, as these contain phase information. Instead, when integral length
scales defined from the two-point correlation of the squared vorticity are used,
the elongation of the initial compact eddy blob was successfully captured. We
have verified that when non-linear dynamics are considered, i.e., when the
Navier-Stokes \cref{eq:ns} are used, both definitions of the integral length
scale increase in time and evidence the elongation of vortices along the axis of
rotation.

Further, note that to prevent the results from being affected by numerical
artifacts, we stopped the simulations when $\lpar$ was about $8$ time smaller
than $2\pi\dpar$. This constraint limited our runs to a duration of $30\,\tf$
($12.7\,\teddy$), and was due to the simulation with $\rossbyeps=0.06$.
Obviously, for the remaining cases, $2\pi\dpar/\lpar>8$ at $t=30\,\tf$. The
decision of when to interrupt the runs were rather arbitrary, but a value
of $8$ for the ratio $2\pi\dpar/\lpar$ is common in DNS of homogeneous
isotropic turbulence \citep{cardesa:2017}.

\section{Scaling Laws for the Energy Dissipation Rate}\label{sec:dissiplaw}

The analysis for the integral length scales in the previous section has
identified two regimes in our data-set. Whereas the group of runs $R1$ display a
dynamics similar to homogeneous isotropic turbulence, runs in the group $R2$ are
characterized by strong anisotropy. In this section, we present results for the
evolution of the energy dissipation rate and seek for similarity relations that
can collapse the data in the different regimes.

\begin{figure}
\captionsetup[subfigure]{labelformat=empty}
\centering
\subfloat[\label{fig:kinetic_energy}]{}
\subfloat[\label{fig:dissipation}]{}
\begin{tikzpicture}
\coordinate (fig1pos) at (0,0);
\coordinate (fig1xlabel) at ([yshift=-2.6cm]fig1pos);
\coordinate (fig1ylabel) at ([xshift=-3.2cm]fig1pos);
\coordinate (caption)    at ([xshift=-3.0cm,yshift=2.25cm]fig1pos);
\coordinate (annotate1)   at (0.85,1.75);
\node[inner sep=0pt] (figleft) at (fig1pos){
      \includegraphics[width=.48\linewidth]
      {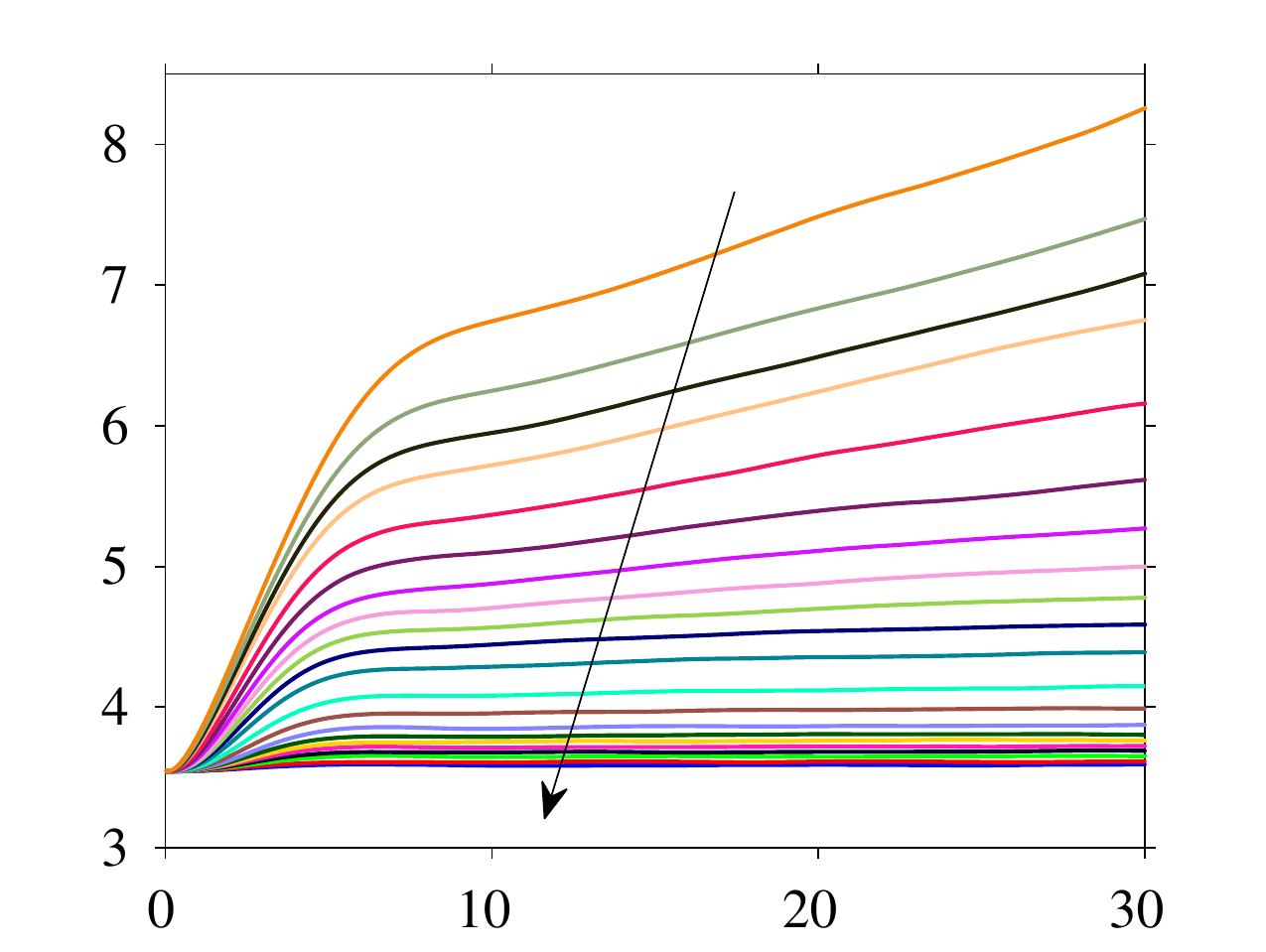}
  };
\node            (xfig1) at (fig1xlabel) {$t/\tf$};
\node[rotate=90] (yfig1) at (fig1ylabel) {$\ken / u_f^2$};
\node            (figa)  at (caption) {(a)};
\node at (annotate1) {$\rossbyeps$};
\coordinate (fig2pos) at (6.4,0);
\coordinate (fig2xlabel) at ([yshift=-2.6cm]fig2pos);
\coordinate (fig2ylabel) at ([xshift=-3.2cm]fig2pos);
\coordinate (caption)    at ([xshift=-3.2cm,yshift=2.25cm]fig2pos);
\coordinate (annotate2)  at (6.25,-0.15);
\node[inner sep=0pt] (figleft) at (fig2pos){
      \includegraphics[width=.48\linewidth]
      {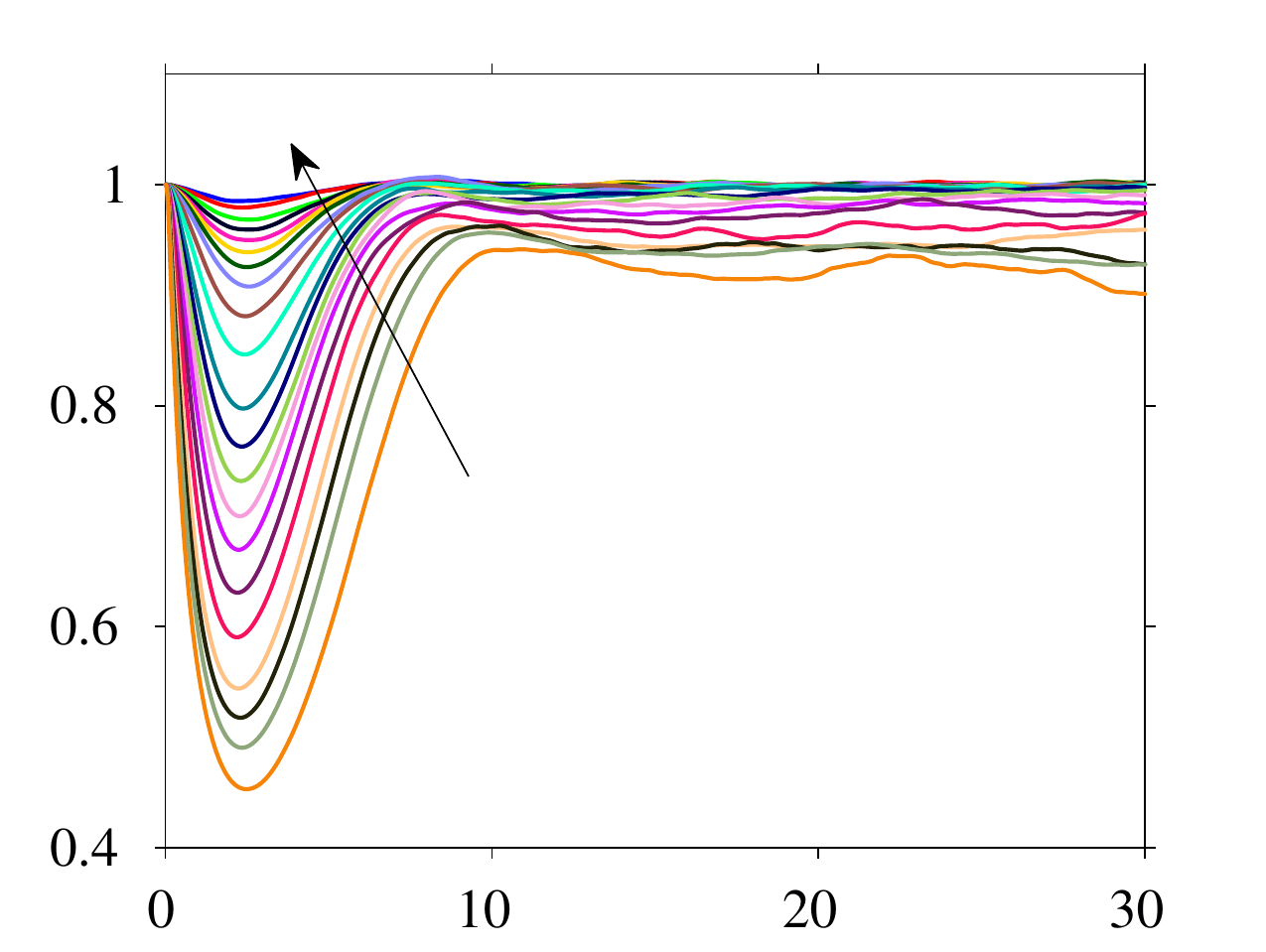}
  };
\node            (xfig2) at (fig2xlabel) {$t/\tf$};
\node[rotate=90] (yfig2) at (fig2ylabel) {$\dissipnu / \dissipI$};
\node            (figb)  at (caption) {(b)};
\node  at (annotate2)   {$\rossbyeps$};
\end{tikzpicture}
 \caption{Time evolution of the box averaged kinetic energy $\ken$ and
energy dissipation rate $\dissipnu$ over time.}
\label{fig:ken_dissip}
\end{figure}
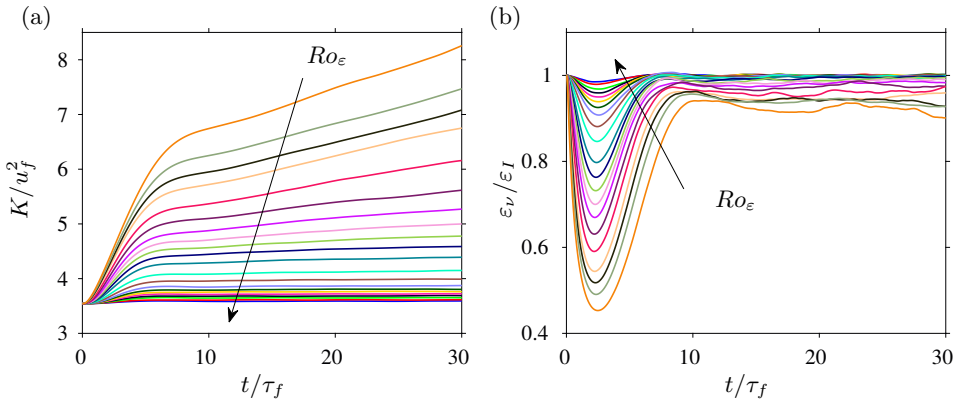

After the onset of rotation, both $\ken$ and $\dissipnu$ evolve in time
according to the conservation of energy, i.e., $\dinline{\ken}{t} = -\dissipnu
+\dissipI$. While $\ken$ grows rapidly (\cref{fig:kinetic_energy}), the viscous
dissipation $\dissipnu$ first decreases monotonically until a minimum that is
inversely proportional to $\rossbyeps$ is reached at roughly $t=3\,\tf$
(\cref{fig:dissipation}). After reaching its lowest value, $\dissipnu$ continues
to grow towards the power input $\dissipI$, although the inequality
$\dissipnu<\dissipI$ remains for some of the cases up to the final simulation
time. Generally speaking, the mismatch between the energy dissipation rate and
the energy input rate in \cref{fig:dissipation} is stronger for smaller
$\rossbyeps$ (group $R2$).

In fact, the imbalance $\dissipnu\ne\dissipI$ is the footprint of an inverse
energy cascade that is triggered by the Coriolis force, and that leads to the
accumulation of energy at the large scales. This is expected to occur when
$\rossbyeps$ is below a critical Rossby number that depends on the geometrical
dimensions of the system \citep{Smith1996a,Deusebio2014,Pestana2019}. In such
cases, however, equilibrium ($\dissipnu=\dissipI)$ can still be restored after
long integration times when the energy in the wavenumbers $\wn<\wnf$ is
sufficiently high to contribute to $\dissipnu$
\citep{Valente2016,Seshasayanan2018}. For the runs considered in this study, the
critical $\rossbyeps$ is approximately $1$ as show in \citet{Pestana2019}.

From \cref{fig:dissipation}, it is evident that a naive scaling in terms of the
forcing parameters can not cause the different lines in \cref{fig:dissipation}
to collapse, as it would in homogeneous isotropic turbulence. In other words, an
approximation of $\dissipnu$ in terms of $\uf$ and $\wnf$ is invalid because the
evolution of $\dissipnu$ in \cref{fig:dissipation} depends clearly on
$\rossbyeps$. In homogeneous isotropic turbulence, the estimation
$\dissipnu\sim\uf^3 \wnf$ suffices since both $\uf$ and $1/\wnf$ are
proportional to a characteristic velocity and a characteristic length, and this
expression is equivalent to $\dissipnu\sim(\urmsiso)^3/\liso$. We must therefore
search for other ways to approximate $\dissipnu$ in rotating turbulence.

\subsection{Spectral Transfer Time}

To address this problem we followed the methodology introduced by
\citet{kraichnan:1965} within the context of MHD and bridged by \citet{zhou1995}
to homogeneous rotating flows. The basic idea is that the rate at which energy
is transferred to the smaller scales depends on an energy content and on a
time scale, viz. the spectral transfer time. If we treat the characteristic
scales as global quantities instead of wavenumber dependent, the dissipation
law can be written in terms of the r.m.s. velocity and the spectral transfer
time as
\begin{equation}
  \dissipnu\sim \frac{\urms^2}{\ts}.
  \label{eq:dissipts}
\end{equation}

The spectral transfer time, however, is composed of two additional time scales,
namely the nonlinear time scale $\tnl$ and the relaxation time scale $\trelax$.
Whereas $\tnl$ indicates how fast the triple velocity correlations are built up
and favors the forward energy cascade, $\trelax$ serves as a relaxation time or
a measure of how fast the triple velocity correlations are destroyed. The
assumptions that the energy dissipation rate $\dissipnu$ is directly
proportional to $\trelax$ and that the energy cascade is local lead to the
so-called ``golden rule'' \citep{zhou1995}:
\begin{equation}
    \ts \sim \frac{\tnl^2}{\trelax}.
\label{eq:ts_definition}
\end{equation}

In \cref{eq:ts_definition}, $\tnl$ involves a velocity and a length scale and
$\trelax$ can rest on any other time scales that are relevant for the problem.
For instance, in forced homogeneous isotropic flows,
$\trelax\sim\tf\sim\tnl\sim\liso/\urmsiso$, which implies
$\ts\sim\liso/\urmsiso$ to recover the well known dissipation law $\dissipnu
\sim (\urmsiso)^3/\liso$, extensively verified by DNS and experiments. For more
complex flows, which involve other time scales like rotating turbulence with the
rotation time scale $\tomg=1/(2\Omega)$, the relaxation time scale $\trelax$ can
be assumed as function of the type $\trelax=\trelax(\tf,\tomg)$
\citep{kraichnan:1965,zhou1995,Matthaeus1989a}. Combining
\cref{eq:dissipts,eq:ts_definition} leads to
\begin{equation}
  \dissipnu\sim \urms^2 \left(\frac{\trelax}{\tnl^2}\right),
\label{eq:dissipdef}
\end{equation}
and the problem of determining the dissipation law becomes the one of
determining $\tnl$ and $\trelax$.

\begin{figure}
\centering
\captionsetup[subfigure]{labelformat=empty}
\subfloat[]{\label{fig:scaling_zh}}
\subfloat[]{\label{fig:scaling_ga}}
\subfloat[]{\label{fig:scaling_cb}}
\subfloat[]{\label{fig:scaling_bq}}
\begin{tikzpicture}
\coordinate (fig1pos) at (0,0);
\coordinate (fig1ylabel) at ([xshift=-5.0cm]fig1pos);
\coordinate (caption)    at ([xshift=-4.7cm,yshift=2.2cm]fig1pos);
\node[inner sep=0pt] (figleft) at (fig1pos){
      \includegraphics[width=.8\linewidth]
      {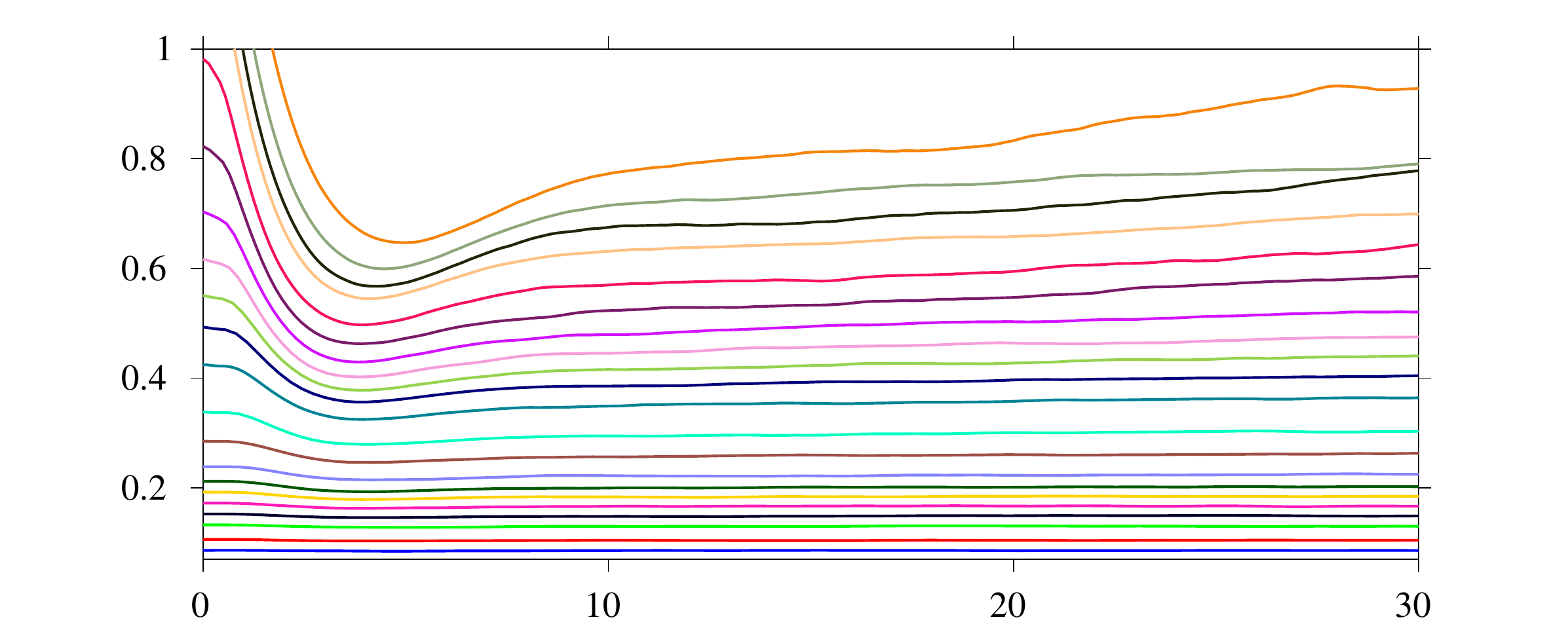}
  };
\node[rotate=90] (yfig1) at (fig1ylabel) {$\dissipnu  \Omega \ell^2 / u'^4$};
\node            (figa)  at (caption) {(a)};
\coordinate (fig2pos) at (0,-4.6);
\coordinate (fig2ylabel) at ([xshift=-5.0cm]fig2pos);
\coordinate (caption)    at ([xshift=-4.7cm,yshift=2.2cm]fig2pos);
\node[inner sep=0pt] (figleft) at (fig2pos){
      \includegraphics[width=.8\linewidth]
      {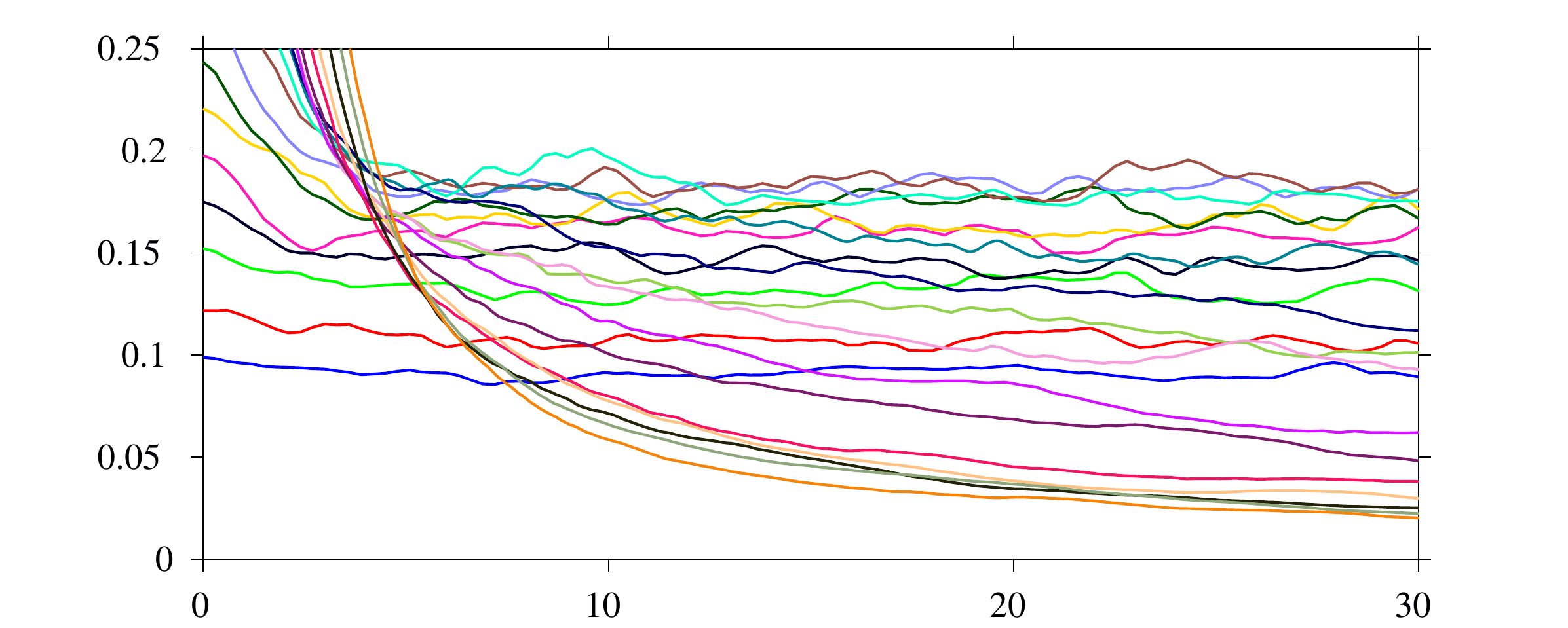}
  };
\node[rotate=90] (yfig2) at (fig2ylabel)
                                    {$\dissipnu  \Omega  \lpp^3 / (u'^4\,\lpar)$};
\node            (figb)  at (caption) {(b)};
\coordinate (fig3pos) at (0,-9.2);
\coordinate (fig3ylabel) at ([xshift=-5.0cm]fig3pos);
\coordinate (caption)    at ([xshift=-4.7cm,yshift=2.2cm]fig3pos);
\node[inner sep=0pt] (figleft) at (fig3pos){
      \includegraphics[width=.8\linewidth]
      {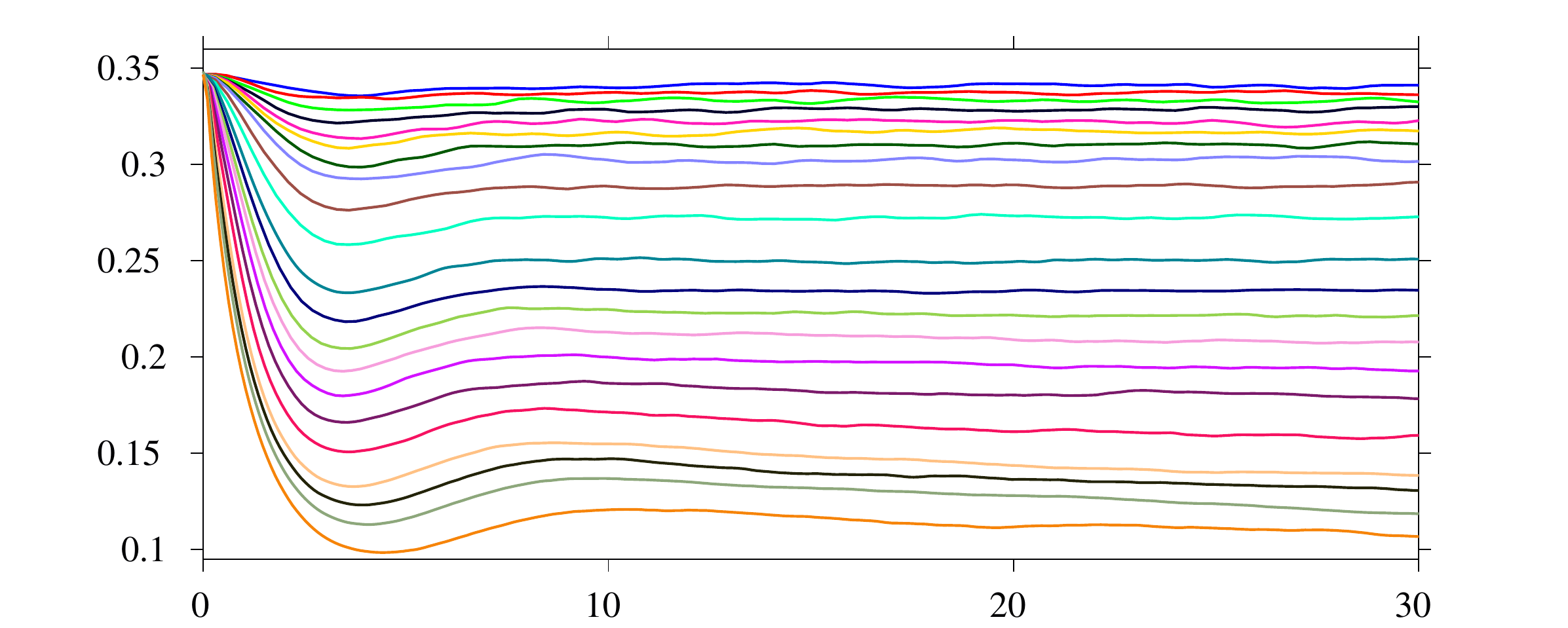}
  };
\node[rotate=90] (yfig3) at (fig3ylabel) {$\dissipnu  \lpp / u'^3$};
\node            (figc)  at (caption) {(c)};
\coordinate (fig4pos) at (0,-13.8);
\coordinate (fig4xlabel) at ([yshift=-2.5cm]fig4pos);
\coordinate (fig4ylabel) at ([xshift=-5.0cm]fig4pos);
\coordinate (caption)    at ([xshift=-4.7cm,yshift=2.2cm]fig4pos);

\node[inner sep=0pt] (figleft) at (fig4pos){
      \includegraphics[width=.8\linewidth]
      {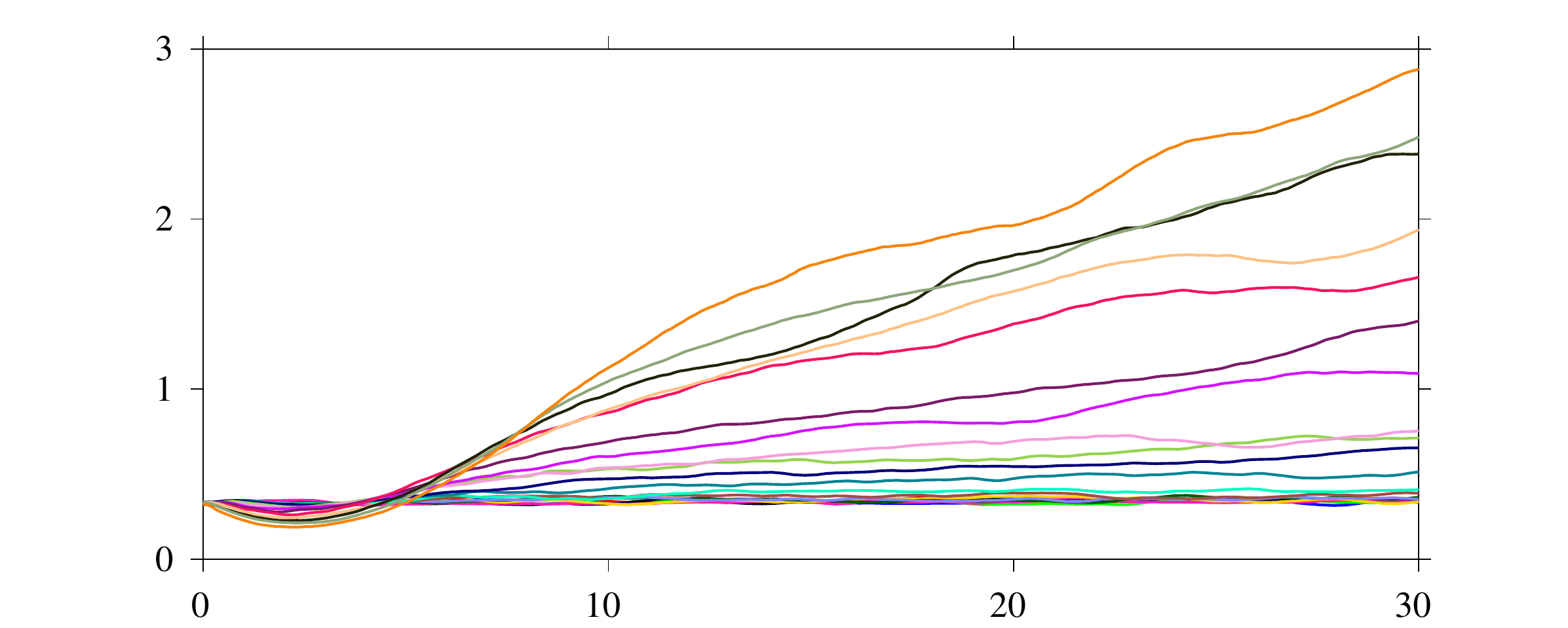}
  };
\node            (xfig4) at (fig4xlabel) {$t/\tf$};
\node[rotate=90] (yfig4) at (fig4ylabel) {$\dissipnu \lpar / u'^3$};
\node            (figd)  at (caption) {(d)};
\draw[-latex] (2,-15.4) -- (2.5,-15.2);
\node at (1.8,-15.35) {\scalebox{0.7}{$R1$}};
\end{tikzpicture}
 \caption{Compensated time evolution of the energy dissipation rate for
$0.06<\rossbyeps<1.54$. Different panels correspond to the different scaling
laws found in the literature: (a) \citet{zhou1995}; (b) Weak inertial-wave
theory \citet{Galtier2003}; (c) Critical balance theory \citet{nazarenko:2011};
(d) \citet{baqui:2015}.}
\label{fig:litscalinglaws}
\end{figure}
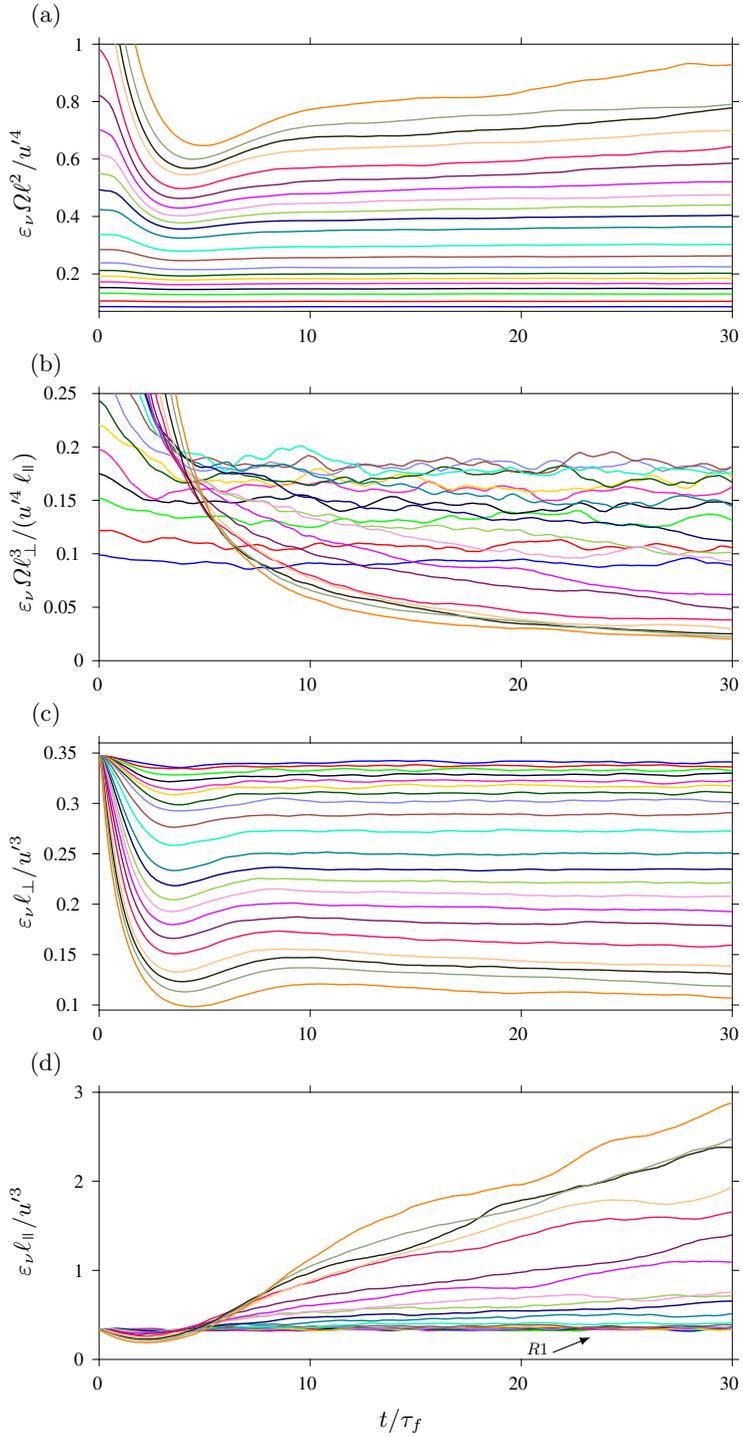

\subsection{Evaluation of Current Available Dissipation Laws}

In current literature, a few dissipation laws for homogeneous rotating
turbulence have been proposed. For example, the approximations that follow from
the theory of \citet{zhou1995,Galtier2003,nazarenko:2011} and \citet{baqui:2015}
are
\begin{equation}
\dissipnu \sim \frac{\urms^4}{\Omega\ell^2}, \quad
\dissipnu \sim \frac{\urms^4 \lpar}{\Omega \lpp^3}, \quad
\dissipnu \sim \frac{\urms^3}{\lpp}, \quad \mbox{and} \quad
\dissipnu \sim \frac{\urms^3}{\lpar}, \label{eq:lit_dissiplaw}
\end{equation}
respectively. Although these authors do not explicitly present their theories
within the framework of a spectral transfer time, we have taken the freedom to
also summarize the theories within this context.

The law proposed by \citet{zhou1995}, for instance, ignores anisotropy. It
assumes that $\tnl\sim\ell/\urms$ and that the relaxation time scale is
proportional to the rotation time scale, i.e., $\trelax\sim\tomg$, to yield
$\dissipnu \sim \urms^4 / (\Omega\ell^2)$. In contrast, dimensional analysis for
the weak inertial-wave theory proposed by \citet{Galtier2003}, which takes into
account scale anisotropy, results in $\dissipnu \sim \urms^4 \lpar/(\Omega
\lpp^3)$, where $\tnl\sim\lpp/\urms$ and $\trelax\sim\lpar/(\Omega\lpp)$. When
anisotropy is however disregarded, i.e., $\ell\sim\lpar\sim\lpp$, the
predictions by \citet{Galtier2003} reduce to the relation proposed by
\citet{zhou1995}. The critical balance theory of \citet{nazarenko:2011}
considers that $\tnl\sim\trelax\sim\lpp/\urms$ and the theory of
\citet{baqui:2015} suggests that $\tnl\sim\trelax\sim\lpar/\urms$.

When we apply the scaling laws in \cref{eq:lit_dissiplaw} to the data presented
in \cref{fig:dissipation}, we observe overall a poor prediction; see
\cref{fig:litscalinglaws}. The approximations of
\cite{zhou1995,Galtier2003,nazarenko:2011} in
\cref{fig:scaling_zh,fig:scaling_ga,fig:scaling_cb} are found insufficient as
the different curves are far away from collapsing into a single curve. For large
$\rossbyeps$, \cref{fig:scaling_zh} delivers at least straight lines, suggesting
a correction factor in terms of $\rossbyeps$. For small $\rossbyeps$, however,
deviations from a straight line are evident. In \cref{fig:scaling_ga}, the
curves of the $5$ last runs in group $R2$ (lowest $\rossbyeps$) seems to follow
a similar trend, yet for high $\rossbyeps$ a poor match is again found.
Differently, \cref{fig:scaling_cb} shows that for large $\rossbyeps$ and
$t>10\,\tf$, the curves are flat and tend closer to another. This is expected as
$\lpp$ must tend to $\liso$ for large $\rossbyeps$, and, in this limit, the
dissipation law of homogeneous isotropic turbulence is recovered.

The best approximation, at least for part of the data-set, is obtained with the
scaling law of \citet{baqui:2015}. \Cref{fig:scaling_bq} indicates that this
scaling is suitable for the runs in group R1 (indicated with an arrow in the
figure). For the other runs in group R2, \cref{fig:scaling_bq} also provides
unsatisfactory results like the other scaling laws. We are then motivated to
look into a similarity law for this group of runs.

\subsection{A Dissipation Scaling Law for Runs in Group R2}

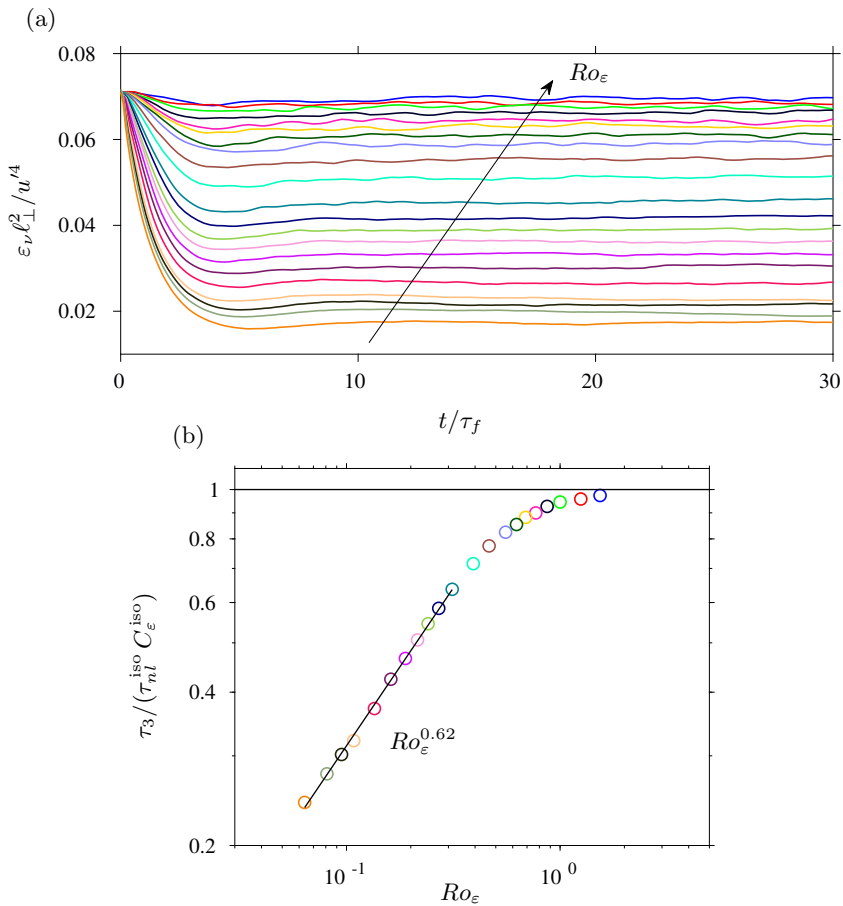
\begin{figure}
\centering
\captionsetup[subfigure]{labelformat=empty}
\subfloat[\label{fig:tau_relax_o_time}]{}
\subfloat[\label{fig:tau_relax_avg}]{}
\begin{tikzpicture}
\coordinate (fig1pos) at (0,0);
\coordinate (fig1xlabel) at ([yshift=-2.8cm]fig1pos);
\coordinate (fig1ylabel) at ([xshift=-5.8cm]fig1pos);
\coordinate (caption)    at ([xshift=-5.55cm,yshift=2.5cm]fig1pos);
\coordinate (annotate)   at (1.7,1.8);
\node[inner sep=0pt] (figleft) at (fig1pos){
      \includegraphics[width=.9\linewidth]
      {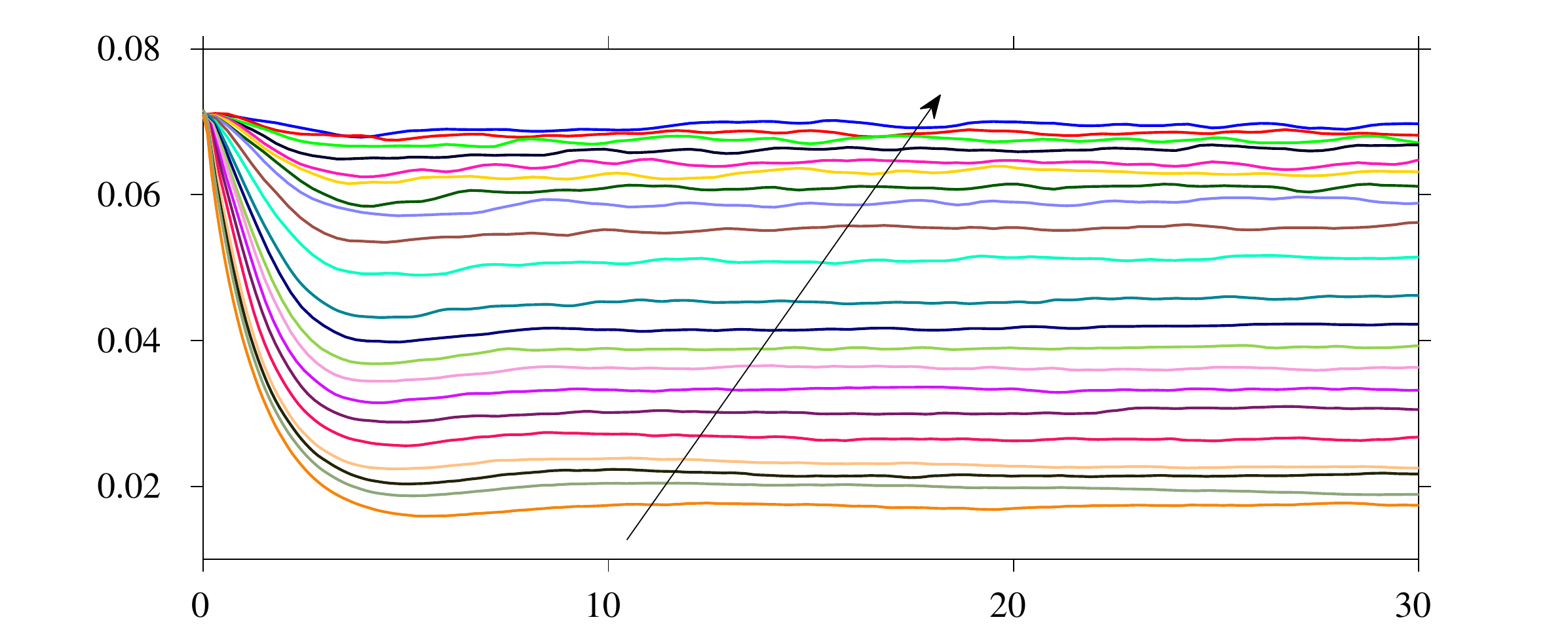}
  };
\node            (xfig1) at (fig1xlabel) {$t/\tf$};
\node[rotate=90] (yfig1) at (fig1ylabel) {$\dissipnu \lpp^2 / \urms^4$};
\node            (figa)  at (caption)    {(a)};
\node at (annotate) {$\rossbyeps$};
\coordinate (fig2pos) at (0,-6.0);
\coordinate (fig2xlabel) at ([yshift=-3.0cm]fig2pos);
\coordinate (fig2ylabel) at ([xshift=-4.2cm]fig2pos);
\coordinate (caption)    at ([xshift=-3.6cm,yshift=-3cm]fig1pos);
\coordinate (annotate)   at (-0.5,-7.0);
\node[inner sep=0pt] (figleft) at (fig2pos){
      \includegraphics[width=.6\linewidth]
      {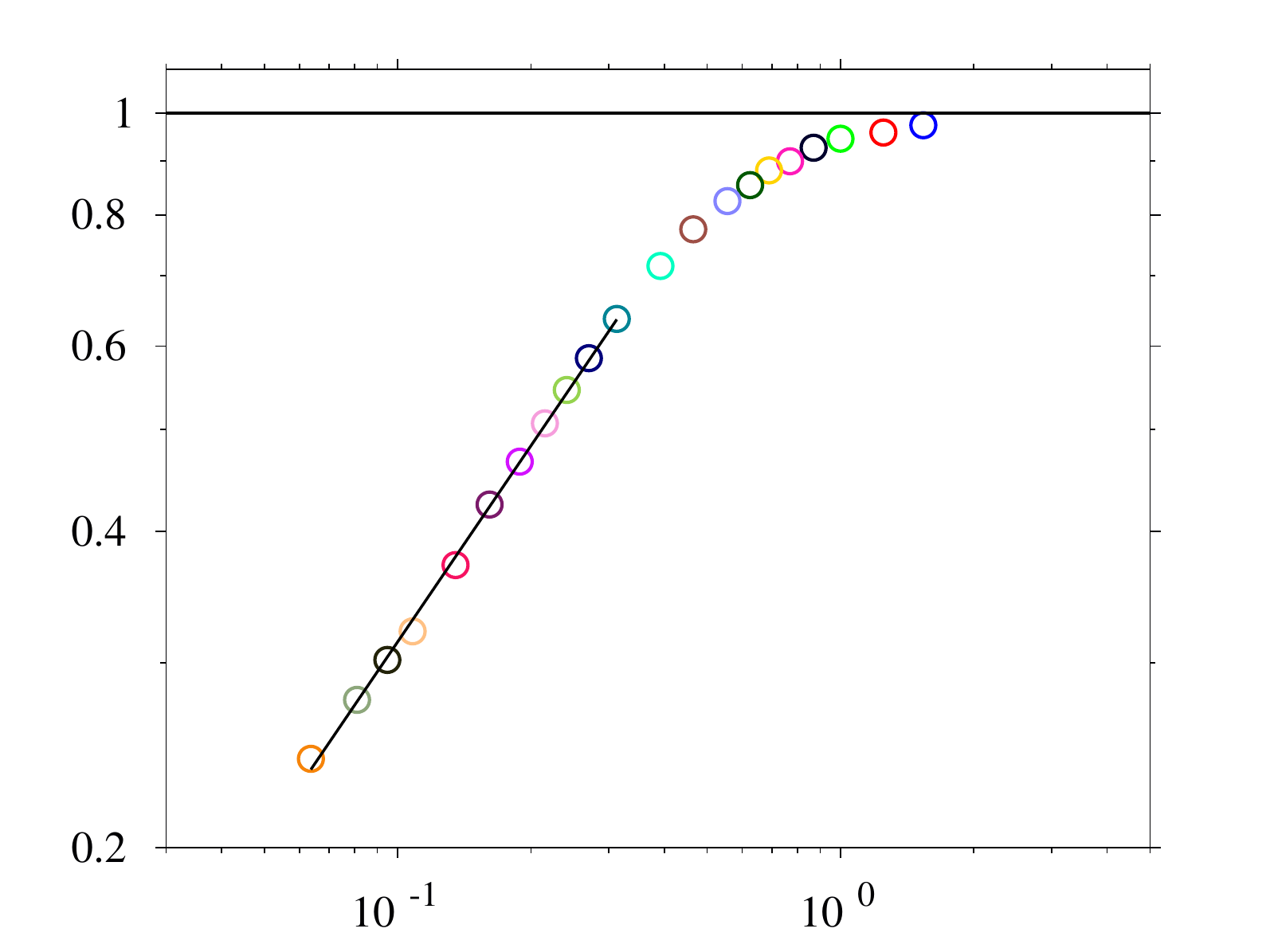}
  };
\node            (xfig2) at (fig2xlabel) {$\rossbyeps$};
\node[rotate=90] (yfig2) at (fig2ylabel) {$\trelax/( \tnliso \, \cepsiso)$};
\node            (figb)  at (caption)    {(b)};
\node at (annotate) {$\rossbyeps^{0.62}$};

\end{tikzpicture}
 \caption{Decorrelation (relaxation) time scale $\trelax$ as function of time
(a) and averaged over the interval $10\le t \le 30\,\tf$ and normalized by the
the non-linear time scale $\tnl$ times the proportionality constant
$C_\varepsilon^{\,\text{iso}}$(b). Two reference lines are included. The
horizontal line at the top signalizes that for large $\rossbyeps$, the
relaxation time scale tends to the value of the non-linear time scale of the
homogeneous isotropic case. The other line shows the power-law dependency of the
type $\rossbyeps^h$ with $h=0.62$ for runs of the group $R2$.}
\label{fig:scaling_law_mod}
\end{figure}

To find a dissipation law for runs in group R2 we base ourselves on
\cref{eq:dissipdef}. The first question we turn to is the one of finding an
approximation for $\tnl$. The non-linear time scale involves an estimation of a
velocity and a length scale, which we shall assume as $\urms$ and $\lpp$,
respectively. The reason behind this choice goes as follow. From
\cref{fig:par_lscale_strong,fig:dissipation}, we observe that $\lpp$ and
$\dissipnu$ display similar dynamics, although inverse; the evolution of each
variable is the opposite of the other. This behavior hints to a dependency of
the form $\dissipnu \sim 1/\lpp$, which can also be justified like in the
critical balance theory \citep{nazarenko:2011}. Within this theory, the basic
idea is that rotation tends to destroy derivatives along the direction of
rotation and the advection term is mainly due to the normal velocity gradients
and the normal velocity field. Thus, $\lpp$ is taken as the relevant
length scale for the non-liner interactions. On the other hand, for the velocity
scale, an alternative would be to take information about the transversal
velocity fields only as in \citet{baqui:2015}. Nevertheless, although rotation
favors two dimensionalization, the velocity field remains three component and
the anisotropy in the Reynolds stress tensor is minimal \citet{Yeung1998}. For
the above reasons, we assume $\tnl \sim \lpp/\urms$, and \cref{eq:dissipdef} can
be in a preliminary step expressed as
\begin{equation}
    \dissipnu \sim \frac{\urms^4}{\lpp^2}  \trelax(\tf,\tomg).
\label{eq:dissip_trelax}
\end{equation}

\begin{figure}
\centering
\begin{tikzpicture}
\coordinate (fig1pos) at (0,0);
\coordinate (fig1xlabel) at ([yshift=-3.0cm]fig1pos);
\coordinate (fig1ylabel) at ([xshift=-4.25cm]fig1pos);
\node[inner sep=0pt] (figleft) at (fig1pos){
      \includegraphics[width=.6\linewidth]
      {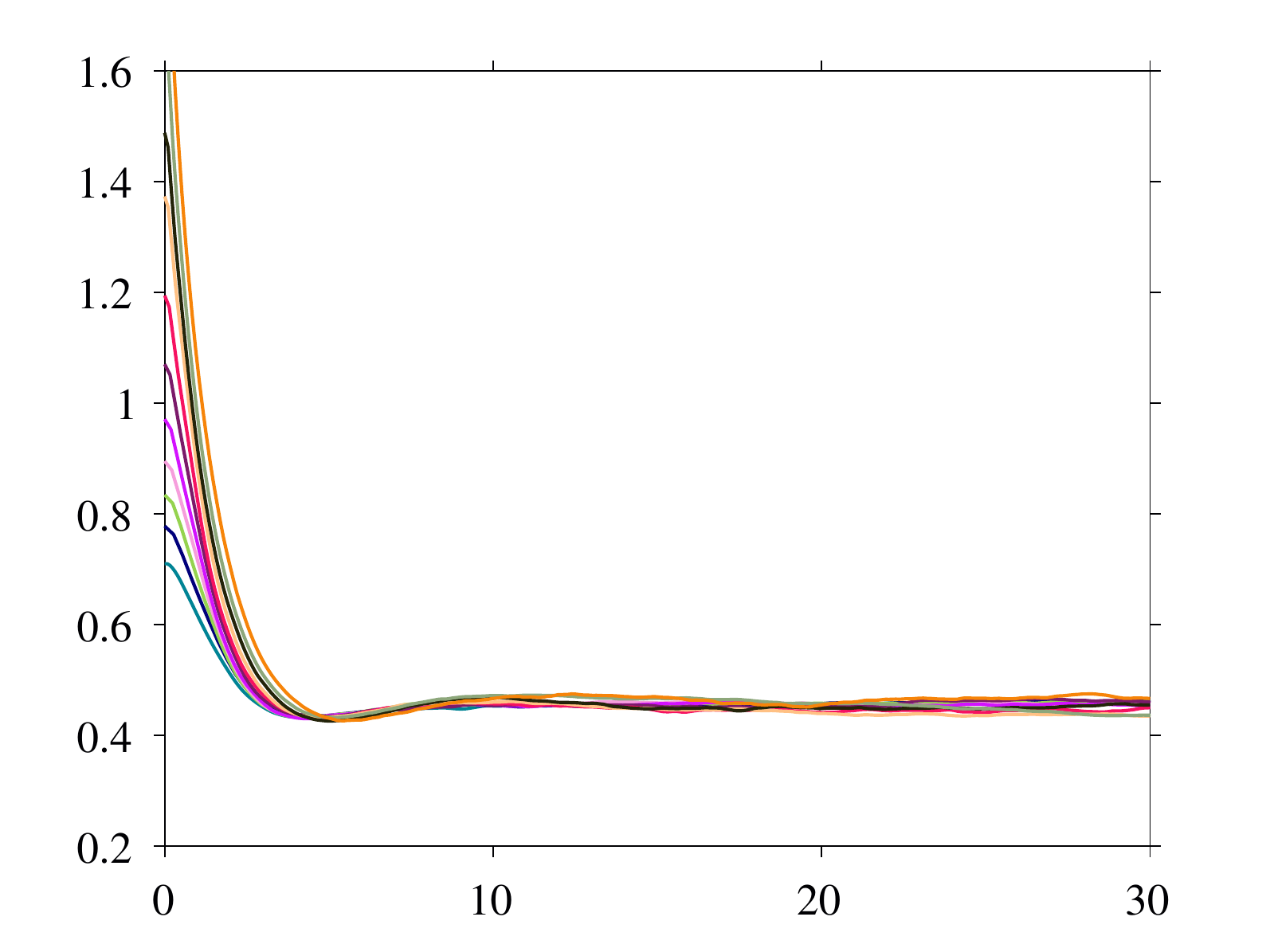}
  };
\node            (xfig1) at (fig1xlabel) {$t/\tf$};
\node[rotate=90] (yfig1) at (fig1ylabel)
                {$\dissipnu \, \lpp^2 / ( \urms^4 \rossbyeps^{0.62} \tnliso)$};
\end{tikzpicture}
 \caption{Time evolution of the energy dissipation rate scaled according to
\cref{eq:dissip_strong_law} for $\rossbyeps\le0.31$.}
\label{fig:dissip_scaled_strong}
\end{figure}

Now, to determine the relaxation time scale, we rearrange
\cref{eq:dissip_trelax} so that $\trelax$ appears as a function of the other
terms and examine its temporal evolution. \Cref{fig:tau_relax_o_time} shows
$\trelax\sim\dissipnu\,\lpp^2 / \urms^4$ over time for all runs. After a
transient of approximately $10\,\tf$, we observe that the curves for different
$\rossbyeps$ reach a plateau, with a terminal value that depends on
$\rossbyeps$.
To determine this dependency, results from \cref{fig:tau_relax_o_time} are then
averaged in the interval $10\,\tf<t<30\,\tf$ and the mean value is shown
against the corresponding Rossby number in \cref{fig:tau_relax_avg}. In the
latter figure, the ordinates appear normalized by the nonlinear time scale
$\tnliso$ times $\cepsiso$, which is the constant of proportionality of the
dissipation law in homogeneous isotropic turbulence. For runs in group $R1$,
$\trelax/(\tnliso \,
\cepsiso)$ increases with $\rossbyeps$ and asymptotically approaches $1$,
implying that for these $\rossbyeps$ the effects of rotation are negligible and
the scaling law of homogeneous isotropic turbulence is recovered. Contrarily and
more surprising, we see that $\trelax/(\tnliso \,\cepsiso)$ follow the
power-law $\rossbyeps^{h}$ with $h=0.62$ for the runs in group $R2$.
Consequently, for this group, we can finally express \cref{eq:dissip_trelax} as
\begin{equation}
    \dissipnu \sim \frac{\urms^4}{\lpp^2} (\tnliso \rossbyeps^{0.62}) \sim
    \frac{\urms^3}{\lpp}
    \left[
    \left(\frac{\tnliso}{\tnl}\right)
    \rossbyeps^{0.62}
    \right].
\label{eq:dissip_strong_law}
\end{equation}

\Cref{eq:dissip_strong_law} summarizes the effects of rotation for the runs with
$0.06<\rossbyeps<0.31$, and suggests that in a rotating frame of reference, the
disparity between $\tnl$ and $\tnliso$ increases, such that the ratio $(\tnl /
\tnliso)$ shrinks with the inverse of $\rossbyeps^{0.62}$. Finally, scaling the
data in \cref{fig:dissipation} with \cref{eq:dissip_strong_law} leads to
\cref{fig:dissip_scaled_strong}, where very good agreement is found for all the
cases in group $R2$.

\section{Conclusions}

We have investigated the effects of system rotation with Rossby numbers in the
range $0.06 \le \rossbyeps \le 1.54$ on the evolution of an initial cloud of
isotropic eddies. Differently from other studies, which have focused on the
initial transient immediately after the onset of rotation, we have focused
instead on longer time intervals. This was only possible because our DNS were
carried out in elongated domains which were $340$ times larger than the initial
characteristic eddy size.

The classical pictures of rotating turbulence were reproduced, in which we
observed the formation of columnar eddies along the axis of rotation and a
decrease in the energy dissipation rate. However, by following the evolution of
the integral length scales we identified different dynamics that were shown to
depend on $\rossbyeps$. This led us to separate our data-set into $2$ groups.
While the runs in group $R1$ did not show any pronounced sign of growth in the
integral length scales, for the runs in group $R2$, $\lpar$ grew substantially
and approximately linearly with time. The latter group of runs can be therefore
associated to a regime where the formation of columns predominate, whereas runs
of group $R1$ are closer to homogeneous isotropic turbulence. Further, we found
that the growth rate of the columnar eddies depends exponentially on
$\rossbyeps$, i.e., $\gamma=a\exp{(b\,\rossbyeps)}$, with $a=4$ and $b=-17$.

The energy dissipation rate in the group of runs $R1$ is well approximated by
the scaling law proposed in \citet{baqui:2015}. For the group $R2$, which
consists of runs at lower $\rossbyeps$, we have shown that the scaling laws
currently available in the literature fail to approximate $\dissipnu$. Still, we
were able to find a similarity relation for $\dissipnu$ in the range
$0.06\le\rossbyeps\le0.31$ by applying the ideas introduced by
\citet{kraichnan:1965}, in which the spectral transfer time is regarded as
composed of two opposing time scales. First, by observing the inverse relation
between $\lpp$ and $\dissipnu$, we assumed that $\lpp$ was the relevant
length scale to form $\tnl$. Second, the relaxation time scale $\trelax$ was
shown to depend on a power-law of $\rossbyeps$ and on $\tnliso$, which implies
that it is exclusively a function of $\rossbyeps$ and of the forcing parameters
$\wnf$ and $\uf$. Thus, we arrived at a similarity law for this $\rossbyeps$
range. Scaling $\dissipnu$ with $\urms^4/(\lpp^2 \rossbyeps^{0.62} \tnliso)$
collapsed the data for different $\rossbyeps$ into a single curve.

Last, we would like to remark that the results for the case where the rotation
rate is highest, i.e., $\rossbyeps=0.06$, were verified by increasing the
numerical resolution and the domain size by a factor $2$ in the direction of
rotation. However, whether other dynamics emerge at even lower $\rossbyeps$ and
the effects of $\reynoldseps$ remains to be studied. In any case, we hope our
numerical investigation provide a useful contribution to the improvement of
turbulence models and stimulates other studies to elucidate and quantify the
effects of the Coriolis force on the evolution of a cloud of isotropic eddies in
unbounded domains. \bibliographystyle{jfm}
\bibliography{manuscript}

\begin{thebibliography}{40}
\expandafter\ifx\csname natexlab\endcsname\relax\def\natexlab#1{#1}\fi
\def\au#1{#1} \def\ed#1{#1} \def\yr#1{#1}\def\at#1{#1}\def\jt#1{\textit{#1}}
  \def\bt#1{#1}\def\bvol#1{\textbf{#1}} \def\vol#1{#1} \def\pg#1{#1}
  \def\publ#1{#1}\def\arxiv#1{#1}\def\org#1{#1}\def\st#1{\textit{#1}}

\bibitem[Alvelius(1999)]{Alvelius1999}
{\sc \au{Alvelius, K.}} \yr{1999}  \at{{Random forcing of three-dimensional
  homogeneous turbulence}}.  \jt{Physics of Fluids}  \bvol{11}~(7),  \pg{1880}.

\bibitem[Baqui \& Davidson(2015)]{baqui:2015}
{\sc \au{Baqui, Yasir~Bin} \& \au{Davidson, P.~a.}} \yr{2015}  \at{{A
  phenomenological theory of rotating turbulence}}.  \jt{Physics of Fluids}
  \bvol{27}~(2),  \pg{025107}.

\bibitem[Bardina {\em et~al.\/}(1985)Bardina, Ferziger \& Rogallo]{Rogallo1985}
{\sc \au{Bardina, Jorge}, \au{Ferziger, J.~H.} \& \au{Rogallo, R.~S.}}
  \yr{1985}  \at{{Effect of rotation on isotropic turbulence: computation and
  modelling}}.  \jt{Journal of Fluid Mechanics}  \bvol{154}~(-1),  \pg{321}.

\bibitem[Bartello {\em et~al.\/}(1994)Bartello, M{\'{e}}tais \&
  Lesieur]{bartello:1994}
{\sc \au{Bartello, Peter}, \au{M{\'{e}}tais, Olivier} \& \au{Lesieur, Marcel}}
  \yr{1994}  \at{{Coherent structures in rotating three-dimensional
  turbulence}}.  \jt{Journal of Fluid Mechanics}  \bvol{273}~(c),  \pg{1--29}.

\bibitem[Batchelor \& Press(1953)]{batchelor:book}
{\sc \au{Batchelor, G~K} \& \au{Press, Cambridge~University}} \yr{1953} {\em
  {The Theory of Homogeneous Turbulence}\/}.  \publ{Cambridge University
  Press}.

\bibitem[Boffetta \& Ecke(2012)]{Boffetta2012}
{\sc \au{Boffetta, Guido} \& \au{Ecke, Robert~E.}} \yr{2012}
  \at{{Two-Dimensional Turbulence}}.  \jt{Annual Review of Fluid Mechanics}
  \bvol{44}~(1),  \pg{427--451}.

\bibitem[van Bokhoven {\em et~al.\/}(2009)van Bokhoven, Clercx, van Heijst \&
  Trieling]{VanBokhoven2009}
{\sc \au{van Bokhoven, L. J.~A.}, \au{Clercx, H. J.~H.}, \au{van Heijst, G.
  J.~F.} \& \au{Trieling, R.~R.}} \yr{2009}  \at{{Experiments on rapidly
  rotating turbulent flows}}.  \jt{Physics of Fluids}  \bvol{21}~(9),
  \pg{096601}.

\bibitem[Cardesa {\em et~al.\/}(2017)Cardesa, Vela-Mart{\'{i}}n \&
  Jim{\'{e}}nez]{cardesa:2017}
{\sc \au{Cardesa, Jos{\'{e}}~I.}, \au{Vela-Mart{\'{i}}n, Alberto} \&
  \au{Jim{\'{e}}nez, Javier}} \yr{2017}  \at{{The turbulent cascade in five
  dimensions}}.  \jt{Science}  \bvol{357}~(6353),  \pg{782--784}.

\bibitem[Delache {\em et~al.\/}(2014)Delache, Cambon \& Godeferd]{Delache2014}
{\sc \au{Delache, Alexandre}, \au{Cambon, Claude} \& \au{Godeferd, Fabien}}
  \yr{2014}  \at{{Scale by scale anisotropy in freely decaying rotating
  turbulence}}.  \jt{Physics of Fluids}  \bvol{26}~(2).

\bibitem[Deusebio {\em et~al.\/}(2014)Deusebio, Boffetta, Lindborg \&
  Musacchio]{Deusebio2014}
{\sc \au{Deusebio, E.}, \au{Boffetta, G.}, \au{Lindborg, E.} \& \au{Musacchio,
  S.}} \yr{2014}  \at{{Dimensional transition in rotating turbulence}}.
  \jt{Physical Review E}  \bvol{90}~(2),  \pg{023005}.

\bibitem[Galtier(2003)]{Galtier2003}
{\sc \au{Galtier, S{\'{e}}bastien}} \yr{2003}  \at{{Weak inertial-wave
  turbulence theory}}.  \jt{Physical Review E}  \bvol{68}~(1),  \pg{015301}.

\bibitem[Godeferd {\em et~al.\/}(2015)Godeferd, Ed \& Moisy]{Godeferd2015}
{\sc \au{Godeferd, Fabien~S.}, \au{Ed, Fr} \& \au{Moisy, Eric}} \yr{2015}
  \at{{Structure and Dynamics of Rotating Turbulence: A Review of Recent
  Experimental and Numerical Results}}.  \jt{Applied Mechanics Reviews}
  \bvol{67}~(3),  \pg{030802}.

\bibitem[Greenspan(1968)]{greenspan1968theory}
{\sc \au{Greenspan, H~P}} \yr{1968} {\em {The Theory of Rotating Fluids}\/}.
  \publ{Cambridge University Press}.

\bibitem[Hopfinger {\em et~al.\/}(1982)Hopfinger, Browand \&
  Gagne]{Hopfinger1982}
{\sc \au{Hopfinger, E.~J.}, \au{Browand, F.~K.} \& \au{Gagne, Y.}} \yr{1982}
  \at{{Turbulence and waves in a rotating tank}}.  \jt{Journal of Fluid
  Mechanics}  \bvol{125}~(-1),  \pg{505}.

\bibitem[Hunt {\em et~al.\/}(1988)Hunt, Wray \& Moin]{Hunt1988}
{\sc \au{Hunt, J. C.~R.}, \au{Wray, A.~A.} \& \au{Moin, P.}} \yr{1988}
  \at{{Eddies, streams, and convergence zones in turbulent flows}}.  \bt{In
  {\em Studying Turbulence Using Numerical Simulation Databases, 2. Proceedings
  of the 1988 Summer Program\/}}, ,  \vol{vol.~1},  \pg{pp. 193--208}.

\bibitem[Ibbetson \& Tritton(1975)]{Ibbetson1975}
{\sc \au{Ibbetson, a.} \& \au{Tritton, D.~J.}} \yr{1975}  \at{{Experiments on
  turbulence in a rotating fluid}}.  \jt{Journal of Fluid Mechanics}
  \bvol{68},  \pg{639}.

\bibitem[Ishihara {\em et~al.\/}(2009)Ishihara, Gotoh \& Kaneda]{Ishihara2009b}
{\sc \au{Ishihara, Takashi}, \au{Gotoh, Toshiyuki} \& \au{Kaneda, Yukio}}
  \yr{2009}  \at{{Study of High–Reynolds Number Isotropic Turbulence by
  Direct Numerical Simulation}}.  \jt{Annual Review of Fluid Mechanics}
  \bvol{41}~(1),  \pg{165--180}.

\bibitem[Jacquin {\em et~al.\/}(1990)Jacquin, Leuchter, Cambon \&
  Mathieu]{Jacquin1990}
{\sc \au{Jacquin, L.}, \au{Leuchter, O.}, \au{Cambon, C.} \& \au{Mathieu, J.}}
  \yr{1990}  \at{{Homogeneous turbulence in the presence of rotation}}.
  \jt{Journal of Fluid Mechanics}  \bvol{220},  \pg{1--52}.

\bibitem[Kaneda {\em et~al.\/}(2003)Kaneda, Ishihara, Yokokawa, Itakura \&
  Uno]{Kaneda2003}
{\sc \au{Kaneda, Yukio}, \au{Ishihara, Takashi}, \au{Yokokawa, Mitsuo},
  \au{Itakura, Ken'ichi} \& \au{Uno, Atsuya}} \yr{2003}  \at{{Energy
  dissipation rate and energy spectrum in high resolution direct numerical
  simulations of turbulence in a periodic box}}.  \jt{Physics of Fluids}
  \bvol{15}~(2).

\bibitem[Kraichnan(1965)]{kraichnan:1965}
{\sc \au{Kraichnan, Robert~H.}} \yr{1965}  \at{{Inertial-Range Spectrum of
  Hydromagnetic Turbulence}}.  \jt{Physics of Fluids}  \bvol{8}~(7),
  \pg{1385}.

\bibitem[Matthaeus \& Zhou(1989)]{Matthaeus1989a}
{\sc \au{Matthaeus, William~H} \& \au{Zhou, Ye}} \yr{1989}  \at{{Extended
  inertial range phenomenology of magnetohydrodynamic turbulence}}.
  \jt{Physics of Fluids B}  \bvol{1}~(9),  \pg{1929--1931}.

\bibitem[Mininni {\em et~al.\/}(2009)Mininni, Alexakis \& Pouquet]{Mininni2009}
{\sc \au{Mininni, P.~D.}, \au{Alexakis, A.} \& \au{Pouquet, A.}} \yr{2009}
  \at{{Scale interactions and scaling laws in rotating flows at moderate Rossby
  numbers and large Reynolds numbers}}.  \jt{Physics of Fluids}  \bvol{21}~(1),
   \pg{015108}.

\bibitem[Mininni {\em et~al.\/}(2012)Mininni, Rosenberg \&
  Pouquet]{mininni:2012}
{\sc \au{Mininni, P.~D.}, \au{Rosenberg, D.} \& \au{Pouquet, A.}} \yr{2012}
  \at{{Isotropization at small scales of rotating helically driven
  turbulence}}.  \jt{Journal of Fluid Mechanics}  \bvol{699}~(1),
  \pg{263--279}.

\bibitem[Moisy {\em et~al.\/}(2011)Moisy, Morize, Rabaud \&
  Sommeria]{Moisy2011}
{\sc \au{Moisy, F.}, \au{Morize, C.}, \au{Rabaud, M.} \& \au{Sommeria, J.}}
  \yr{2011}  \at{{Decay laws, anisotropy and cyclone-anticyclone asymmetry in
  decaying rotating turbulence}}.  \jt{Journal of Fluid Mechanics}  \bvol{666},
   \pg{5--35}.

\bibitem[Morinishi {\em et~al.\/}(2001)Morinishi, Nakabayashi \&
  Ren]{Morinishi2001a}
{\sc \au{Morinishi, Y.}, \au{Nakabayashi, K.} \& \au{Ren, S.~Q.}} \yr{2001}
  \at{{New DNS algorithm for rotating homogeneous decaying turbulence}}.
  \jt{International Journal of Heat and Fluid Flow}  \bvol{22}~(1),
  \pg{30--38}.

\bibitem[Nazarenko \& Schekochihin(2011)]{nazarenko:2011}
{\sc \au{Nazarenko, Sergei~V.} \& \au{Schekochihin, Akexander~A.}} \yr{2011}
  \at{{Critical balance in magnetohydrodynamic, rotating and stratified
  turbulence: towards a universal scaling conjecture}}.  \jt{Journal of Fluid
  Mechanics}  \bvol{677},  \pg{134--153},  \arxiv{arXiv: 0904.3488}.

\bibitem[Pekurovsky(2012)]{p3dfft}
{\sc \au{Pekurovsky, Dmitry}} \yr{2012}  \at{{P3DFFT: A Framework for Parallel
  Computations of Fourier Transforms in Three Dimensions}}.  \jt{SIAM Journal
  on Scientific Computing}  \bvol{34}~(4),  \pg{C192--C209}.

\bibitem[Pestana \& Hickel(2019)]{Pestana2019}
{\sc \au{Pestana, Tiago} \& \au{Hickel, Stefan}} \yr{2019}  \at{{Regime
  transition in the energy cascade of rotating turbulence}}.  \jt{Physical
  Review E}  \bvol{99}~(5),  \pg{053103}.

\bibitem[Pope(2000)]{pope_turbulent_flows}
{\sc \au{Pope, S~B}} \yr{2000} {\em {Turbulent Flows}\/}.  \publ{Cambridge
  University Press}.

\bibitem[Rogallo(1977)]{Rogallo1977}
{\sc \au{Rogallo, R.~S.}} \yr{1977}  \at{{An ILLIAC program for the numerical
  simulation of homogeneous incompressible turbulence}}.  \jt{NASA Technical
  Memo}  \bvol{73},  \pg{203}.

\bibitem[Seshasayanan \& Alexakis(2018)]{Seshasayanan2018}
{\sc \au{Seshasayanan, Kannabiran} \& \au{Alexakis, Alexandros}} \yr{2018}
  \at{{Condensates in rotating turbulent flows}}.  \jt{Journal of Fluid
  Mechanics}  \bvol{841},  \pg{434--462}.

\bibitem[Smith {\em et~al.\/}(1996)Smith, Chasnov \& Waleffe]{Smith1996a}
{\sc \au{Smith, Leslie~M.}, \au{Chasnov, Jeffrey~R.} \& \au{Waleffe, Fabian}}
  \yr{1996}  \at{{Crossover from Two- to Three-Dimensional Turbulence}}.
  \jt{Physical Review Letters}  \bvol{77}~(12),  \pg{2467--2470}.

\bibitem[Staplehurst {\em et~al.\/}(2008)Staplehurst, Davidson \&
  Dalziel]{Staplehurst2008}
{\sc \au{Staplehurst, P~J}, \au{Davidson, P~A} \& \au{Dalziel, S~B}} \yr{2008}
  \at{{Structure formation in homogeneous freely decaying rotating
  turbulence}}.  \jt{Journal of Fluid Mechanics}  \bvol{598},  \pg{81--105}.

\bibitem[Tang {\em et~al.\/}(2018)Tang, Antonia, Djenidi, Danaila \&
  Zhou]{Tang2018}
{\sc \au{Tang, S.~L.}, \au{Antonia, R.~A.}, \au{Djenidi, L.}, \au{Danaila, L.}
  \& \au{Zhou, Y.}} \yr{2018}  \at{{Reappraisal of the velocity derivative
  flatness factor in various turbulent flows}}.  \jt{Journal of Fluid
  Mechanics}  \bvol{847},  \pg{244--265}.

\bibitem[Valente \& Dallas(2016)]{Valente2016}
{\sc \au{Valente, Pedro~C.} \& \au{Dallas, Vassilios}} \yr{2016}  \at{{Spectral
  imbalance in the inertial range dynamics of decaying rotating turbulence}}
  \bvol{023114},  \pg{1--8},  \arxiv{arXiv: 1610.05032}.

\bibitem[{Van Atta} \& Antonia(1980)]{VanAtta1980}
{\sc \au{{Van Atta}, C.~W.} \& \au{Antonia, R.~A.}} \yr{1980}  \at{{Reynolds
  number dependence of skewness and flatness factors of turbulent velocity
  derivatives}}.  \jt{Physics of Fluids}  \bvol{23}~(2),  \pg{252--257}.

\bibitem[Yeung \& Zhou(1998)]{Yeung1998}
{\sc \au{Yeung, P.~K.} \& \au{Zhou, Ye}} \yr{1998}  \at{{Numerical study of
  rotating turbulence with external forcing}}.  \jt{Physics of Fluids}
  \bvol{10}~(11),  \pg{2895}.

\bibitem[Yoshimatsu {\em et~al.\/}(2011)Yoshimatsu, Midorikawa \&
  Kaneda]{Yoshimatsu2011}
{\sc \au{Yoshimatsu, K.}, \au{Midorikawa, M.} \& \au{Kaneda, Y.}} \yr{2011}
  \at{{Columnar eddy formation in freely decaying homogeneous rotating
  turbulence}}.  \jt{Journal of Fluid Mechanics}  \bvol{677},  \pg{154--178}.

\bibitem[Zeman(1994)]{Zeman1994}
{\sc \au{Zeman, O.}} \yr{1994}  \at{{A note on the spectra and decay of
  rotating homogeneous turbulence}}.  \jt{Physics of Fluids}  \bvol{6}~(10),
  \pg{3221}.

\bibitem[Zhou(1995)]{zhou1995}
{\sc \au{Zhou, Ye}} \yr{1995}  \at{{A phenomenological treatment of rotating
  turbulence}}.  \jt{Physics of Fluids}  \bvol{7}~(8),  \pg{2092}.

\end{thebibliography}

\end{document}